\documentclass[aps,reprint,amsmath,amssymb,superscriptaddress]{revtex4-2}

\usepackage{amsmath}
\usepackage{amssymb}
\usepackage{graphicx}
\usepackage{CJK}
\usepackage{keyval}
\usepackage{dcolumn}
\usepackage{bm}
\usepackage{color}
\usepackage{txfonts}
\usepackage{verbatim}
\usepackage[rightcaption]{sidecap}

\usepackage{soul}

\begin{document}

\title{Orientation-dependent electron-phonon coupling in interfacial superconductors LaAlO$_3$/KTaO$_3$}

\author{Xiaoyang Chen}
\affiliation{State Key Laboratory of Surface Physics, Department of Physics, and Advanced Materials Laboratory, Fudan University, Shanghai 200438, China}

\author{Tianlun Yu}
\affiliation{State Key Laboratory of Surface Physics, Department of Physics, and Advanced Materials Laboratory, Fudan University, Shanghai 200438, China}

\author{Yuan Liu}
\affiliation{School of Physics, Zhejiang University, Hangzhou, 310027, China}

\author{Yanqiu Sun}
\affiliation{School of Physics, Zhejiang University, Hangzhou, 310027, China}

\author{Minyinan Lei}
\affiliation{State Key Laboratory of Surface Physics, Department of Physics, and Advanced Materials Laboratory, Fudan University, Shanghai 200438, China}

\author{Nan Guo}
\affiliation{State Key Laboratory of Surface Physics, Department of Physics, and Advanced Materials Laboratory, Fudan University, Shanghai 200438, China}

\author{Yu Fan}
\affiliation{State Key Laboratory of Surface Physics, Department of Physics, and Advanced Materials Laboratory, Fudan University, Shanghai 200438, China}

\author{Xingtian Sun}
\affiliation{State Key Laboratory of Surface Physics, Department of Physics, and Advanced Materials Laboratory, Fudan University, Shanghai 200438, China}

\author{Meng Zhang}
\affiliation{School of Physics, Zhejiang University, Hangzhou, 310027, China}

\author{Fatima Alarab}
\affiliation{Swiss Light Source, Paul Scherrer Institute, Villigen, CH-5232, Switzerland}

\author{Vladimir N. Strokov}
\affiliation{Swiss Light Source, Paul Scherrer Institute, Villigen, CH-5232, Switzerland}

\author{Yilin Wang}
\affiliation{School of Future Technology and Department of Physics, University of Science and Technology of China, Hefei, 230026, China}

\author{Tao Zhou}
\affiliation{State Key Laboratory of Surface Physics, Department of Physics, and Advanced Materials Laboratory, Fudan University, Shanghai 200438, China}

\author{Xinyi Liu}
\affiliation{State Key Laboratory of Surface Physics, Department of Physics, and Advanced Materials Laboratory, Fudan University, Shanghai 200438, China}

\author{Fanjin Lu}
\affiliation{State Key Laboratory of Surface Physics, Department of Physics, and Advanced Materials Laboratory, Fudan University, Shanghai 200438, China}

\author{Weitao Liu}
\affiliation{State Key Laboratory of Surface Physics, Department of Physics, and Advanced Materials Laboratory, Fudan University, Shanghai 200438, China}

\author{Yanwu Xie}
\email{ywxie@zju.edu.cn}
\affiliation{School of Physics, Zhejiang University, Hangzhou, 310027, China}

\author{Rui Peng}
\email{pengrui@fudan.edu.cn}
\affiliation{State Key Laboratory of Surface Physics, Department of Physics, and Advanced Materials Laboratory, Fudan University, Shanghai 200438, China}
\affiliation{Shanghai Research Center for Quantum Sciences, Shanghai 201315, China}

\author{Haichao Xu}
\email{xuhaichao@fudan.edu.cn}
\affiliation{State Key Laboratory of Surface Physics, Department of Physics, and Advanced Materials Laboratory, Fudan University, Shanghai 200438, China}
\affiliation{Shanghai Research Center for Quantum Sciences, Shanghai 201315, China}

\author{Donglai Feng}
\email{dlfeng@ustc.edu.cn}
\affiliation{School of Future Technology and Department of Physics, University of Science and Technology of China, Hefei, 230026, China}
\affiliation{Shanghai Research Center for Quantum Sciences, Shanghai 201315, China}
\affiliation{Collaborative Innovation Center of Advanced Microstructures, Nanjing 210093, China}

\date{\today}% It is always \today, today,
             %  but any date may be explicitly specified

\begin{abstract}
The emergent superconductivity at the LaAlO$_3$/KTaO$_3$ interfaces exhibits a mysterious dependence on the KTaO$_3$ crystallographic orientations. Here we show, by soft X-ray angle-resolved photoemission spectroscopy, that the interfacial superconductivity is contributed by mobile electrons with unexpected quasi-three-dimensional character, beyond the “two-dimensional electron gas” scenario in describing oxide interfaces. At differently-oriented interfaces, the quasi-three-dimensional electron gas ubiquitously exists and spatially overlaps with the small $q$ Fuchs-Kliewer surface phonons. Intriguingly, electrons and the Fuchs-Kliewer phonons couple with different strengths depending on the interfacial orientations, and the stronger coupling correlates with the higher superconducting transition temperature. Our results provide a natural explanation for the orientation-dependent superconductivity, and the first evidence that interfacial orientations can affect electron-phonon coupling strength over several  nanometers, which may have profound implications for the applications of oxide interfaces in general.
\end{abstract}

\maketitle

\section{\label{sec:level1}Introduction}
With the great success of semiconductor interfaces in electronic and photonic applications over the past 50 years, interfaces between complex oxides bring new hope for next-generation multifunctional device applications. One paradigm example is the LaAlO$_3$/SrTiO$_3$ (LAO/STO) interface, where the ``two-dimensional electron gas" (2DEG) emerges at the interface between two band-insulators \cite{r1}. and further becomes superconducting at the transition temperature ($T_{\mathrm{c}})\sim $200~mK \cite{r2}. 
So far its superconducting paring mechanism remains debated \cite{r3, r4, r5, r6, r7, r8, r9, r10, r11, r12, r13}.
Recently, a second family of oxide interfacial superconductors is discovered at LaAlO$_3$/KTaO$_3$ (LAO/KTO) and EuO/KTaO$_3$, which soon becomes a new research spotlight \cite{r14, r15, r16, r17, r18, r19, r20, r21}. Remarkably, the superconductivity develops at $T_{\mathrm{c}}\sim$ 2~K at LAO/KTO(111) \cite{r14, r15}, $T_{\mathrm{c}}\sim$ 0.9~K at LAO/KTO(110) \cite{r19}, but is absent at LAO/KTO(001) down to 25~mK \cite{r14}.
The higher optimal-$T_{\mathrm{c}}$ than LAO/STO and the extraordinary orientation-dependent superconductivity in KTO-based interfaces offer a new perspective for exploring the characters and mechanism of interfacial superconductivity between oxide insulators. 

The orientation-dependent superconductivity has never been reported in any other superconductors, whose origin remains a tantalizing puzzle. 
There are several possible explanations. 
First, if some of the mobile electrons are confined to a single-interfacial layer \cite{r22, r23, r24} and are crucial to the superconductivity, the electron-phonon coupling (EPC) and interfacial superconductivity could be sensitive to the local atomic configuration at the interface \cite{r25, r26, r27}.
However, this scenario seems inconsistent with the estimated superconducting layer thickness over 4~nm based on upper critical field measurements of the KTO-based interfaces \cite{r14, r15}. 
Secondly, pairing through inter-orbital interactions mediated by soft transverse optical (TO) phonons has been proposed to explain the superconductivity \cite{r21, r28}, where the orientation dependence is attributed to different orbital configurations due to dimensional confinement \cite{r21}. 
In contrast to the three degenerate $t_{2g}$ orbitals in the interfacial states of LAO/KTO(111), it is proposed that the degeneracy of orbitals is reduced to two and one in LAO/KTO(110) and LAO/KTO(001), respectively, which could suppress the inter-orbital hopping and superconductivity \cite{r21}. 
Thirdly, the coupling between electrons and longitudinal optical (LO) phonons has been proposed to mediate superconducting pairing in LAO/STO \cite{r8, r9, r11, r12, r13}, however, it is unclear whether/how it can cause orientation-dependent superconductivity.
To examine the existing scenarios, direct measurements on the interfacial electronic states, orbital characters and EPC at differently oriented KTO-based interfaces are demanded.

In the superconducting LAO/KTO heterostructures, the mobile electrons are generally buried below the insulating LAO layers of over 10 nm thickness \cite{r15, r19}. Only recently it is discovered that 1.5 nm-AlO$_x$/1 nm-LAO/KTO(111) retains superconductivity with relative thinner overlayer \cite{r29}. It is still challenging for angle-resolved photoemission spectroscopy (ARPES) measurements due to its surface sensitivity. Though the bulk-sensitive hard X-ray ARPES could reach the buried interfacial states \cite{r18}, it lacks the energy/momentum resolution for clearly revealing the dispersive information. Here we overcome this difficulty by studying superconducting interfaces with the thinnest overlayers and exploiting $\sim$1000~eV-photon-excited soft X-ray (SX-) ARPES with adequate probing depth and decent energy/momentum resolution \cite{r30, r31}. Our results demonstrate orientation dependence of the coupling strength between the interfacial mobile electrons and interfacial modes of high energy optical phonons of KTO, which strikingly correlate with the superconductivity.

\section{Results}
\subsection{Transport properties}
Amorphous overlayers were grown on top of KTO with (111), (110) and (001) orientations [Fig. 1(a)] using pulsed laser deposition \cite{r29}. 
With the optimized overlayer of 1.5 nm-AlO$_x$/1 nm-LAO/KTO, the temperature dependence of the sheet resistance ($R_\mathrm{sheet}$) shows superconducting transitions with $T_\mathrm{c}^\mathrm{middle}$ = 1.22 K and 0.35 K for (111) and (110) orientations, respectively [Fig. 1(b)]. As for the (001) orientation, superconducting transition is not observed down to 0.1~K. Two-terminal resistances of 3 nm LAO/KTO show $T_\mathrm{c}^\mathrm{middle}$ = 1.3~K, 0.7~K and  $\textless$ 0.4~K for LAO/KTO(111), LAO/KTO(110) and LAO/KTO(001), respectively [Fig. S1]. Although the $T_\mathrm{c}$ is slightly lower than LAO/KTO with thicker LAO \cite{r15, r19}, the orientation dependence of $T_\mathrm{c}$(111) $> T_\mathrm{c}$(110) $> T_\mathrm{c}$(001) persists, in both AlO$_x$/LAO/KTO and LAO/KTO. The two-dimensional carrier density ($n_{2\mathrm{D}}$) at 9~K was extracted from field-dependent Hall resistance  [Fig.~1(c)], as summarized in Fig.~1(d).
According to the superconducting phase diagrams on electrical gate tuned EuO/KTO(111) \cite{r16} and EuO/KTO(110) \cite{r21}, 
our samples locate at the optimal $n_{2\mathrm{D}}$ region.  
The electron mobility ($\mu$) at 9~K is determined from the corresponding $n_{2\mathrm{D}}$ and $R_\mathrm{sheet}$, showing $\mu$(111)~\textless~$\mu$(110)~\textless~$\mu$(001) [Fig.~1(d)], whose trend is consistent with previous reports on LAO/KTO with  a thicker overlayer \cite{r15, r17, r19, r32} and will be discussed later. These transport properties indicate that our samples of different orientations host the typical characters of LAO/KTO interfaces despite of the reduced thickness of the surface insulating overlayers.

\begin{figure}[t]
\includegraphics[width=86mm]{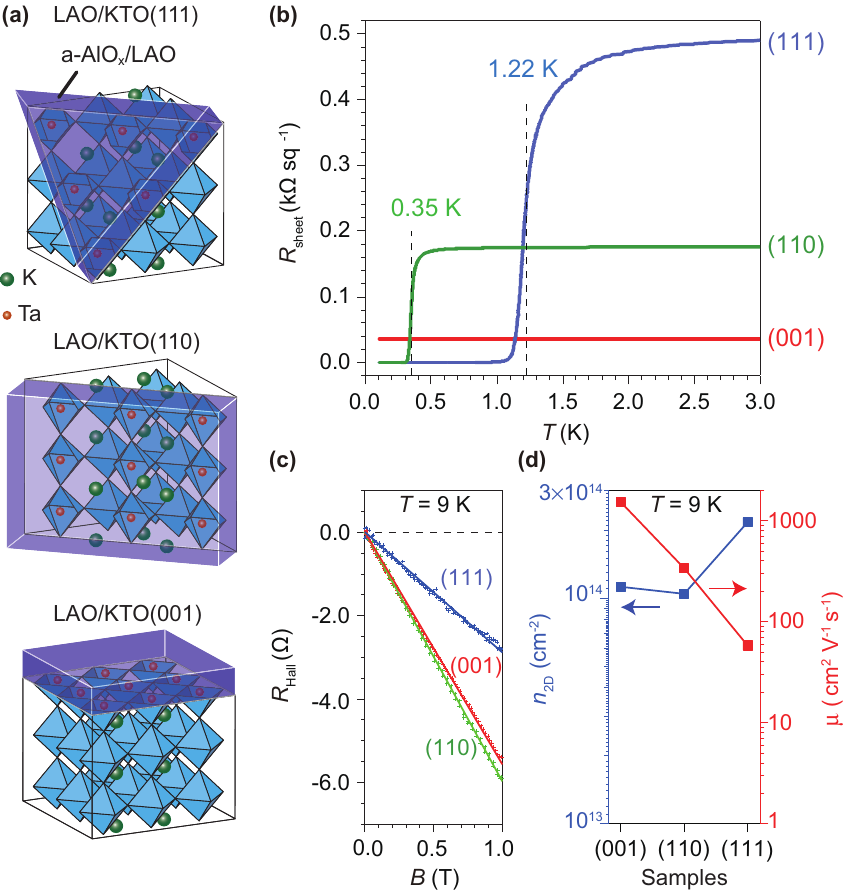}
\caption{\textbf{Orientation-dependent superconductivity of LAO/KTO interfaces.} \textbf{(a)} Sketch of LAO/KTO interfaces with (111), (110), and (001) orientations. \textbf{(b)} Temperature-dependent $R_\mathrm{sheet}$ at the LAO/KTO interfaces with (111), (110), and (001) orientations. The $T_\mathrm{c}^\mathrm{middle}$ is determined by $R_\mathrm{sheet}( T_\mathrm{c}^\mathrm{middle}) = 0.5 \times R_\mathrm{sheet}$(3K). The lowest temperature for measurement is 0.1~K. 
\textbf{(c)} Field-dependent $R_\mathrm{Hall}$ of the same set of samples at 9~K. \textbf{(d)} Two-dimensional carrier density $n_{2\mathrm{D}}$ and carrier mobility $\mu$ determined at the same set of samples at 9~K. $n_{2\mathrm{D}}$ is extracted from a linear fitting to the data in (c). $\mu$ is extracted from the data in (b) and (c), with $\mu^{-1}=R_\mathrm{sheet}n_\mathrm{2D}e$, where $e$ is the elementary charge.
}
\label{fig1}
\end{figure}

\begin{figure*}[t]
\includegraphics[width=170mm]{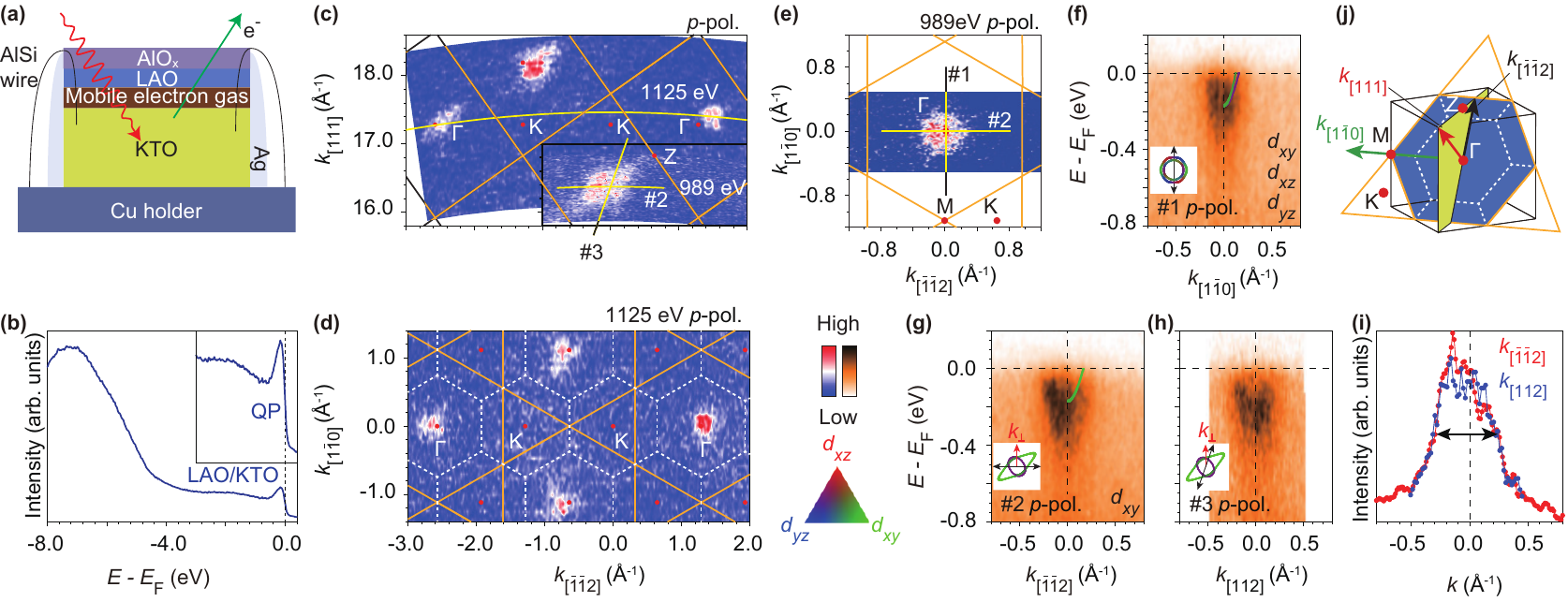}
\caption{\textbf{Quasi-three-dimensional character of the mobile electronic states at the LAO/KTO(111) interface.}
\textbf{(a)} Sketch of the sample mounting and grounding for SX-ARPES measurements. 
\textbf{(b)} Integrated energy distribution curve (EDC) of LAO/KTO(111). The inset shows the zoomed-in quasi-particle (QP) peak of interfacial states.
\textbf{(c)} Out-of-plane photoemission intensity map integrated over [$E_\mathrm{F}-150$~meV, $E_\mathrm{F}+150$~meV] in  $k_{[\bar{1}\bar{1}2]}$-$k_{[111]}$ plane using $p$-polarized ($p$-pol.) photons from 921~eV to 1280~eV. 
\textbf{(d-e)} In-plane photoemission intensity maps integrated over [$E_\mathrm{F}-150$~meV, $E_\mathrm{F}+150$~meV] using 1125~eV  and 989~eV photons, respectively. The corresponding $k_{\perp}$ locations are marked in (c).
The solid orange lines in (d) and (j) illustrate the bulk Brillouin zones, while the white dashed lines illustrate the surface Brillouin zones.
\textbf{(f-h)} Photoemission spectra along \#1 ($k_{[1\bar{1}0]}$), \#2 ($k_{[\bar{1}\bar{1}2]}$) and \#3 ($k_{[112]}$), respectively. (h) is generated from the photon energy dependent data in (c). The corresponding momentum locations are marked in (c) and (e). The calculated bulk KTO bands are overlaid on the right side after a chemical potential shift to match the experimental data, whose orbital characters are noted (see details in Fig. S6). The insets show the calculated Fermi surfaces in (c) and (e). 
\textbf{(i)} Momentum distribution curves (MDCs) integrated over [$E_\mathrm{F}-100$ meV, $E_\mathrm{F}+100$ meV] of data in (g) and (h). 
\textbf{(j)} Sketch of KTO bulk Brillouin zone, zone boundaries from truncation along (111) plane (orange solid line), and the surface Brillouin zone (white dashed line). Some related high symmetric directions are indicated by arrows.
}
\label{fig2}
\end{figure*}

\subsection{Dimensionality of the interfacial states}
As shown in Fig. 2(a), the interfaces were well grounded by AlSi-wire and silver paste for SX-ARPES measurements.
The angle integrated energy distribution curve (EDC) of LAO/KTO(111)
shows a peak near $E_\mathrm{F}$ [Fig. 2(b)], indicating metallic interfacial states in contrast to its insulating components.
The peak near $E_\mathrm{F}$ has been always present since the beginning of measurements, and it shows little variation during photon irradiation [Fig. S2], distinct from the irradiation-induced metallic states in KTO surfaces \cite{r33, r34, r35, r36}, reflecting the intrinsic mobile electronic states at the LAO/KTO interfaces. 
The density of states (DOS) at $E_\mathrm{F}$ is prominently contributed by dispersive features [Figs.~2(c)-2(i)], suggesting that the interfacial mobile electrons accumulate at the crystalline KTO side rather than at amorphous LAO side.

\begin{figure*}
\includegraphics[width=120mm]{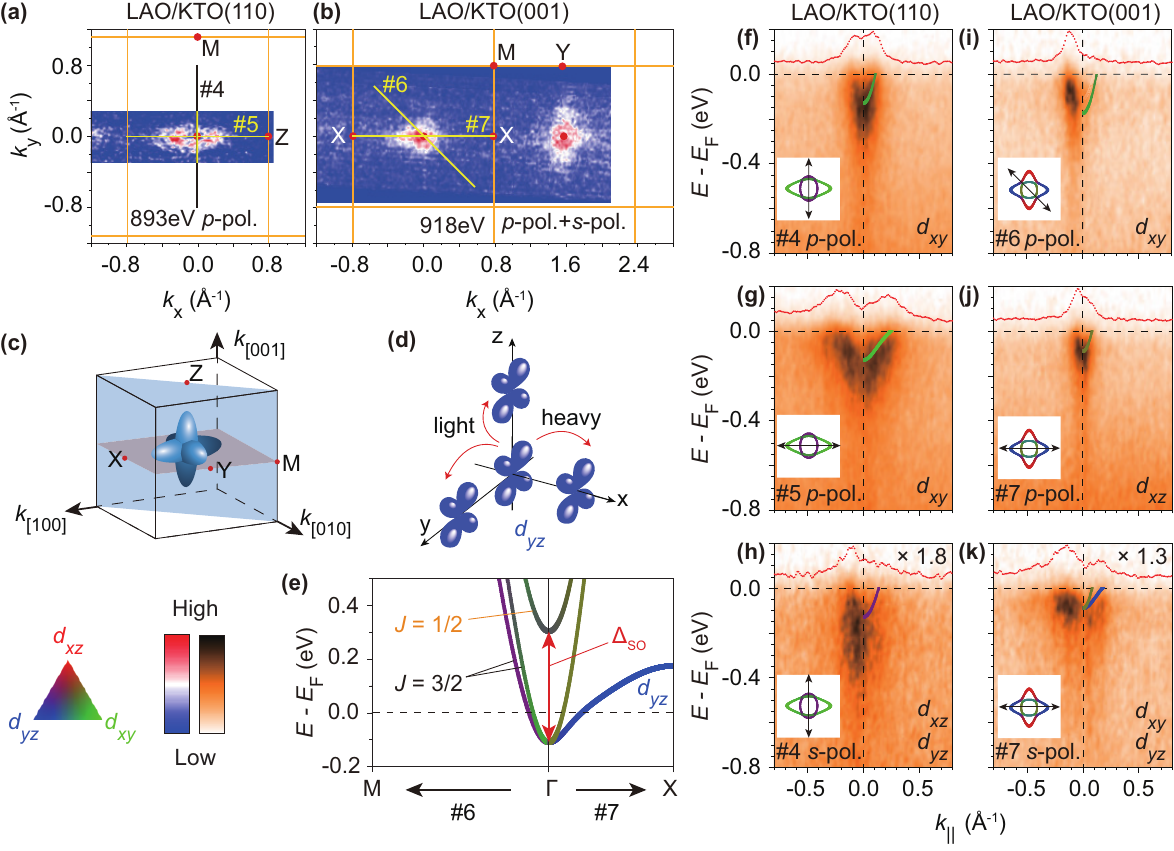}
\caption{\textbf{Interfacial electronic structure of LAO/KTO(110) and LAO/KTO(001).}
\textbf{(a-b)} In-plane photoemission intensity map across $\Gamma$ of LAO/KTO(110) and LAO/KTO(001), respectively. The intensity is integrated over [$E_\mathrm{F}-150$~meV, $E_\mathrm{F}+150$~meV]. 
\textbf{(c)} Sketch of the ellipsoid-like Fermi surfaces of electron-doped KTO and truncating planes (110) and (001). 
\textbf{(d)} Sketch of the electron hopping between neighboring $d_{yz}$ orbitals.
\textbf{(e)} Calculated band dispersion of electron-doped KTO along M-$\Gamma$-X. The spin-orbital coupling (SOC) in KTO mixes the three $t_{2g}$ orbitals and lifts $J=1/2$ band up by approximately 0.4~eV. It leaves two $J=3/2$ bands crossing the $E_\mathrm{F}$ to form the double-layer Fermi surfaces, while the extremal parts of the extended lobes retain almost single orbital character of the corresponding $t_{2g}$ orbital, as shown for $d_{yz}$. 
\textbf{(f-k)} Photoemission spectra along cuts \#4, \#5, \#6 and \#7 in (a-b). 
MDCs integrated between [$E_\mathrm{F}-35$~meV, $E_\mathrm{F}+35$~meV] are overlaid at the top of each panel.
The calculated dispersions of the observed bands are overlaid on the right of each panel, whose orbital characters are noted (see details in Fig. S5-S8).
The insets show the calculated Fermi surfaces and the corresponding cut direction as a double-head arrow. Data from $s$-polarized geometry, which are strongly suppressed owing to matrix element effect \cite{r45}, are amplified by a factor in (h) and (k).
}
\label{fig3}
\end{figure*}

Intriguingly, in stark contrast to a pure two-dimensional states, the measured interfacial electronic structure is highly dispersive along $k_\perp$ ($k_{[111]}$ for LAO/KTO(111) in Fig.~2(c)). 
Thanks to the improved $k_\perp$ resolution of SX-ARPES \cite{r37}, the elongated shape of Fermi surfaces is well resolved in the out-of-plane photoemission mappings [Fig. 2(c)].
The in-plane Fermi surfaces also follow the periodicity of the bulk Brillouin zones (orange solid lines in Fig.~2(d)) rather than the two-dimensional Brillouin zones of the surface atomic layer (white dashed lines in Fig.~2(d)). 
Along two equivalent cuts in the cubic Brillouin zone, the in-plane $k_{[\bar{1}\bar{1}2]}$ (\#2, Fig.~2(g)) and the dominantly out-of-plane $k_{[112]}$ (\#3, Fig.~2(h)), the photoemission spectra show similar features. 
The momentum distribution curves (MDCs) at $E_\mathrm{F}$ are also identical in width [Fig.~2(i)]. Such accordance indicates that the out-of-plane momentum broadening, which combines the effects from the thickness confinement of interfacial state and photoemission detection depth, is comparable to the in-plane momentum resolution, further demonstrating the quasi-three-dimensionality of the electronic states. 
Based on the Fermi surface volume (see Section 7 of Supplemental Material), the carrier density ($n_{3\mathrm{D}}$) was extracted by Luttinger theorem \cite{r38}. Considering possible inhomogeneity and phase separation with coexisting insulating domains at oxide interfaces \cite{r39, r40}, the $n_{3\mathrm{D}}$ from ARPES should represent the domains of higher carrier dopings. Thus the lower thickness limit of the mobile electrons ($d_e$) can be estimated quantitatively by $d_e=n_{2\mathrm{D}}/n_{3\mathrm{D}}$, where $n_{2\mathrm{D}}$ is determined by Hall resistance measurements.
Based on the fine maps of Fermi surfaces and high-statistic scans [Figs. 2(e)-2(g)], $n_{3\mathrm{D}}$ is estimated to be ~3.2~$\times~10^{20}$~cm$^{-3}$ for LAO/KTO(111) interfacial state (see Section 7 of Supplemental Material). Therefore, $d_e$ is estimated to be 6.8~nm for LAO/KTO(111). We also resolved the interfacial electronic states in LAO/KTO(110) and LAO/KTO(001) samples, and all the interfacial states show quasi-three-dimensional character [Fig. S4]. The same analysis conducted gives $d_e$ as 6.0~nm and 5.5~nm for LAO/KTO(110) and LAO/KTO(001), respectively (see Section 7 of Supplemental Material), which are at the same scale as that of LAO/KTO(111). 

The distribution of mobile electrons over the thickness more than 5.5$\sim$6.8~nm indicates that the interfacial mobile electrons are all spatially extended into KTO side over 14$\sim$17~unit cells. 
The superconducting layer thickness ($d_{\mathrm{SC}}$) and the superconducting coherence length ($\xi$) can be estimated by upper critical fields based on the Ginzburg-Landau theory \cite{r41}. For the same sample of LAO/KTO(111), the perpendicular and parallel upper critical field are measured, giving $\xi_{\parallel}\sim$~20~nm and $d_{\mathrm{SC}}\sim$~5~nm [Fig. S3]. 
The superconducting layer thickness roughly agrees with the electron gas thickness determined by ARPES.
The larger superconducting coherence length than the superconducting thickness is consistent with previous studies that suggest two-dimensional character of superconductivity at the LAO/KTO(111) interfaces \cite{r14, r15}.
As the two-dimensional superconductivity is contributed by quasi-three-dimensional electronic states, the critical ingredients that lead to the orientation dependent superconductivity should be in action for more than 5 nanometers. 

\subsection{Orbital composition of the interfacial states}
Orbital composition of the electronic states can be compared at differently-orientated interfaces to scrutinize the orbital-related pairing scenario \cite{r21}
The photoemission intensity maps of LAO/KTO(110), LAO/KTO(001) and LAO/KTO(111) all show Fermi surfaces extending along the $\Gamma$-X, $\Gamma$-Y and $\Gamma$-Z directions [Figs.~3(a)-3(b), 2(c)]. 
The elongated Fermi surface lobes at three perpendicular directions agree well with the expected Fermi surface sheets of electronic bands formed by Ta $t_{2g}$ orbitals ($d_{xy}$, $d_{xz}$, $d_{yz}$) [Fig.~3(c)] \cite{r42}. 
As depicted in Fig.~3(d) for the $d_{yz}$ orbital, the overlap with neighboring $d_{yz}$ is smaller along $x$ direction than those along $y$ and $z$ directions, resulting in a heavy band mass  and a larger $k_\mathrm{F}$ along $\Gamma$-X direction as shown by the theoretical calculations [Fig. 3(e)].  Similarly, $d_{xz}$ and $d_{xy}$ orbitals show heavy band mass and elongation of Fermi surfaces along $\Gamma$-Y and $\Gamma$-Z directions, respectively. 

The polarization-dependent ARPES measurements is a powerful tool to identify orbital characters \cite{r43, r44} (the observable orbitals are noted on the right part in Figs.~3(f)-3(k), 2(f), 2(g), see analysis details in Section 5 of Supplemental Material). 
Combining the data from both $p$-polarized and $s$-polarized geometries, all the three $t_{2g}$  orbitals are identified in LAO/KTO(110) [Figs.~3(f)-3(h)] and in LAO/KTO(001) [Figs.~3(i)-3(k)], with similar electron occupations as those of LAO/KTO(111) [Figs.~2(f), 2(g)]. Except for some fine structure differences (see Section 6 of Supplemental Material for details), the electronic structure and orbital characters basically agree with density functional theory calculations considering spin-orbital coupling on bulk KTO upon a chemical potential shift  [Figs.~3(f)-3(k), 2(f), 2(g)].
These experimental observations exclude a strong difference in orbital occupation numbers among differently-orientated LAO/KTO interfaces, disfavoring the direct relation between orientation-dependent superconductivity and orbital occupations \cite{r21}.

\begin{figure}[htbp]
\centering
\includegraphics[width=86mm]{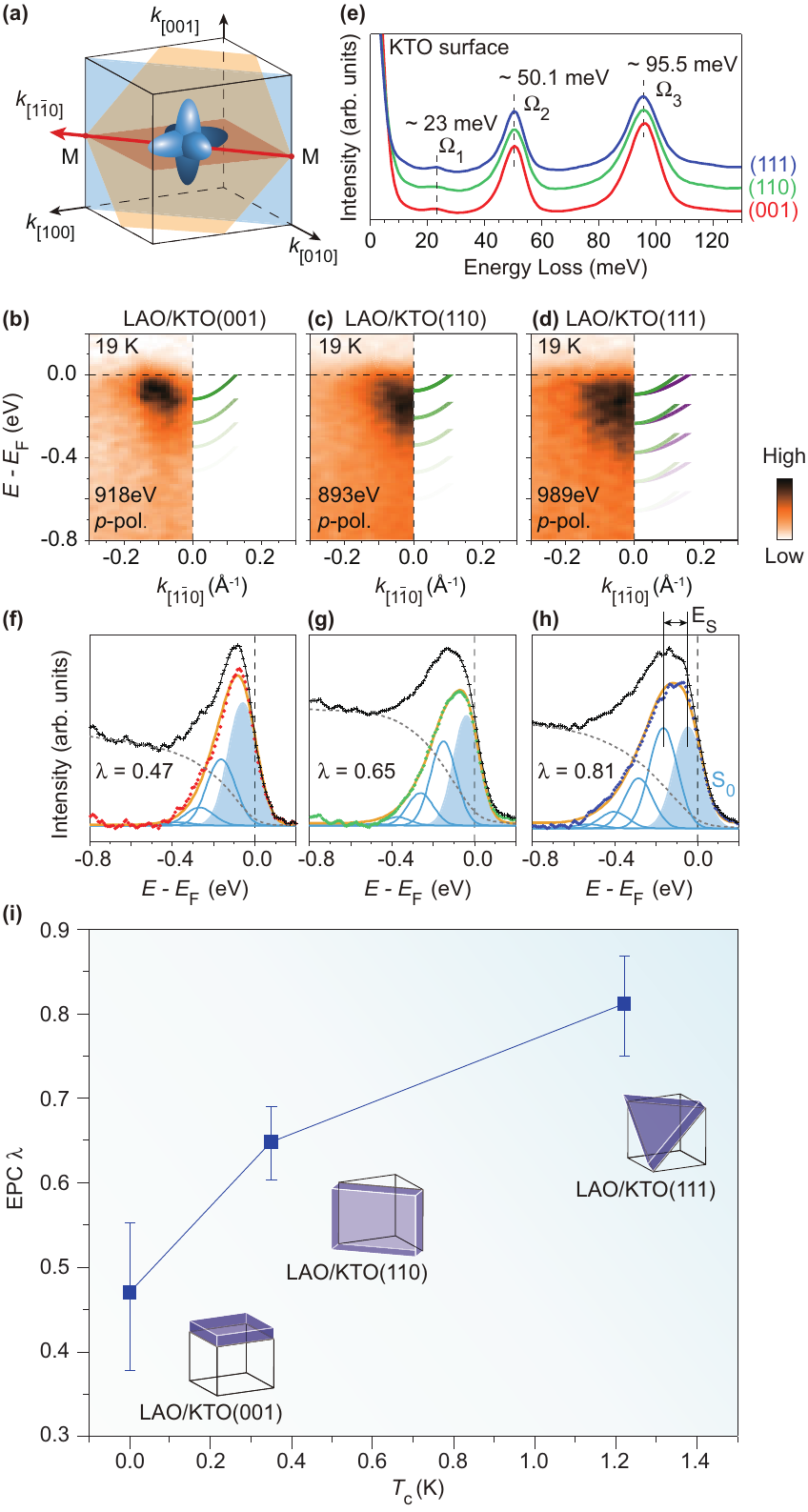}
\caption{\textbf{Orientation-dependent electron-phonon coupling in LAO/KTO interfaces.}
\textbf{(a)} Sketch of the in-plane BZs of LAO/KTO(111) (orange), LAO/KTO(110) (blue), and LAO/KTO(001) (red). 
The cut along $k_{[1\bar{1}0]}$ ($\Gamma$-M) is marked by a red arrow.
\textbf{(b-d)} Photoemission spectra along $k_{[1\bar{1}0]}$ 
for LAO/KTO(001), LAO/KTO(110), and LAO/KTO(111), respectively. Data were measured at 19 K.
The renormalized main band and replicas are illustrated on the right side of each panel, whose transparency reflects the relative intensity as the fitting results in (f-h).
\textbf{(e)} High resolution electron energy loss spectra of KTO(001), KTO(110), and KTO(111) surfaces. Three phonon modes can be identified.
\textbf{(f-h)} EDCs integrated between [$-0.3~\AA^{-1}$, $0.3~\AA^{-1}$] of photoemission spectra (b-d).  The EDCs are fitted by Franck-Condon model after subtracting Tougaard backgrounds (grey dashed curve), which are commonly used for inelastically scattered electrons \cite{r48, r49, r50} [Fig. S11]. The bandwidth and energy separation of replicas are renormalized by the EPC \cite{r51} (see Section 8 of Supplemental Material for details).
(\textbf{i}) EPC coupling constant $\lambda$ obtained from the fittings in (f-h).
}
\label{fig4}
\end{figure}

\subsection{Orientation-dependent EPC}
The high symmetry direction M-$\Gamma$-M ($k_{[1\bar{1}0]}$) in the bulk Brillouin zone is shared as an equivalent in-plane momentum cut for three interfaces with (111), (110) and (001) orientations [Fig.~4(a)], along which the photoemission spectra of three samples are compared with calculations to explore the origin of the orientation-dependent superconductivity [Figs.~4(b)-4(d)]. 
By adjusting the chemical potential of the calculated band structure to match the $k_\mathrm{F}$ of our photoemission data, the bottom of $t_{2g}$ conduction band should locate at binding energy $E_\mathrm{B}\simeq$~0.17~eV for LAO/KTO(111).
However, the photoemission spectral weight extends far beyond the calculated band bottom [Figs.~4(b)-4(d)], with a long tail down to $E_\mathrm{B}\sim$~0.4~eV for LAO/KTO(111) [Fig. 4(d)]. 
Since the spectral tails at higher binding energies retain the momentum distribution of bands at Fermi energy, they are not likely the secondary electrons of random scattering processes. 
Electron correlation is not likely a major cause either, as it usually reduces the bandwidth rather than enlarges it \cite{r46, r47}.
Such a spectral-tail behavior with retained momentum range is a typical polaronic behavior of electrons interacting with the optical phonons at small $q$ region near the zone center \cite{r7, r8, r9}.
Intriguingly, LAO/KTO(111) has the most prominent spectra tail [Fig. 4(d)], followed by LAO/KTO(110) [Fig. 4(c)], and LAO/KTO(001) the least [Fig. 4(b)], suggesting orientation-dependent polaronic behavior and EPC strength. The stronger of the EPC, the higher superconducting $T_\mathrm{c}$.

Superconductivity is achieved at KTO surfaces under electric gating, showing similar $T_{\mathrm{c}}$ and orientation dependence as the LAO/KTO interfaces \cite{r52, r53}.
In contrast, superconductivity is absent in chemically doped bulk KTO \cite{r54, r55}, indicating that the surface phonons of KTO,  rather than bulk KTO or LAO phonons, are likely responsible for the superconductivity of KTO-related surfaces and interfaces. 
High resolution electron energy loss spectra (HR-EELS) measurements were performed to characterize the phonon modes at KTO surface.
As shown in Fig.~4(e), three phonon modes with  energies $\Omega_1\sim 23.2$ meV, $\Omega_2 \sim 50.1$ meV and $\Omega_3 \sim 95.5$ meV are observed at the KTO(111), KTO(110) and KTO(001) surfaces. 
$\Omega_2$ and $\Omega_3$ are slightly lower in energy than the corresponding bulk LO modes ($\Omega_\mathrm{LO3}=52$ meV and $\Omega_\mathrm{LO4}=103$ meV) \cite{r56, r57, r58, r59}, indicating that they are the corresponding LO-derived surface Fuchs-Kliewer modes \cite{r60, r61}. 
It is known that Fuchs-Kliewer modes decay exponentially into the bulk by $e^ {-qd}$, where $d$ is the distance away from the surface or interface and $q$ is the wave vector \cite{r62}. 
For small $q$, the Fuchs-Kliewer modes at the LAO/KTO interface would extend deep into the bulk of KTO. 
Decay length over 20 nm was reported in the Fuchs-Kliewer modes of polar semiconductors \cite{r63}. 
Such depth scale of the small $q$ Fuchs-Kliewer modes of LAO/KTO interface overlap spatially with the nanometer thick interfacial electronic states, which allows the EPC that generates the polaronic behavior. 

According to the Franck-Condon model, the relative intensity among different replica bands should follow Poisson distribution \cite{r8, r9, r64}. As a semi-quantitative estimation on the EPC, we fit the momentum-integrated EDCs as a function of energy ($\varepsilon=E-E_\mathrm{F}$) and the dimension-less EPC coupling constant ($\lambda$):
$$I(\varepsilon, \lambda)= \sum_{l=0}^{n}\frac{[\alpha(\lambda)]^{l}}{l!} S_l(\varepsilon, \lambda)$$
where $\alpha(\lambda)$ is related with EPC \cite{r51}. The DOS of the main band is represented by a resolution-convolved peak $S_0(\varepsilon, \lambda)$ (see details in Section 8 of Supplemental Material) with its bandwidth ($W$) renormalized by EPC as $W=W_0/(1+\lambda)$ \cite{r65}. $S_l(\varepsilon, \lambda)$ represents the DOS of the $l$-th replica, which follows the lineshape of the main band but has an energy shift of $l\times E_\mathrm{S}(\lambda)$. $E_\mathrm{S}(\lambda)$ is the energy separation between replicas, which is slightly larger than the phonon energy due to a renormalization by EPC \cite{r51}. Both phonons with $\Omega_2$ and those with $\Omega_3$ could interact with electrons, while for simplicity, only phonons with energy $\Omega_3$ are included in the fitting, which generates good fitting results for the experimental momentum-integrated spectra (Figs. 4(f)-4(h), see details in Section 8 of Supplemental Material). 
The relative spectra weight of the main band (blue shade in Figs.~4(f)-4(h)), determined by $Z=\int S_\mathrm{0}(\varepsilon)d\varepsilon~\bigg/\int I(\varepsilon)d\varepsilon$, are 0.58, 0.46 and 0.37 for LAO/KTO(001), LAO/KTO(110) and LAO/KTO(111), respectively.
The orientation dependence $Z(111)<Z(110)<Z(001)$ is consistent with that of the mobility $\mu(111)<\mu(110)<\mu(001)$ [Fig. 1(d)].  
The estimated $\lambda$ by such fittings are 0.47, 0.65, and 0.81 for LAO/KTO(001), LAO/KTO(110), and LAO/KTO(111), respectively. 
The EPC coupling strength with $\lambda$(111) $>\lambda$(110) $>\lambda$(001) coincides with that of the superconductivity with $T_\mathrm{c}$(111) $>T_\mathrm{c}$(110) $>T_\mathrm{c}$(001) from both the qualitative comparison and the semi-quantitative fittings.

\section{Summary and Outlook}
Our results demonstrate that the charge transfer, electron doping and orbital characters are similar at differently-orientated LAO/KTO interfaces, and  the tuning parameter of superconductivity in LAO/KTO is likely the coupling between interfacial mobile electrons and the surface Fuchs-Kliewer phonons of KTO with small $q$. 
Phonon mediated superconducting pairing in LAO/KTO is consistent with the recent superfluid stiffness measurements suggesting a nodeless superconducting order parameter in AlO$_x$/KTO(111) interface \cite{r20}. 
Surface Fuchs-Kliewer phonons could be sensitive to sample surfaces, while the detailed variation among crystalline orientations and how it induces the observed orientation-dependent electron-phonon coupling still call for future theoretical development. 
The measured Fermi momenta and electron-phonon coupling behavior in our study can provide experimental foundations for constructing theories describing electrons coupling with surface/interface Fuchs-Kliewer phonons and the orientation dependent superconductivity.
Furthermore, our observation demonstrates that interfacial orientations can affect electron phonon coupling strength over several nanometers, which could provide new routes for engineering various functional properties that are closely related with electron-phonon coupling, including ferroelectricity, multiferroism, and superconductivity.

\begin{acknowledgments}
This work is supported in part by the National Natural Science Foundation of China (Grants Nos. 11790312, 11922403, 11888101, 12074074, 12274085, 11934016), Shanghai Rising-Star Program (Grant No. 20QA1401400) and Shanghai Municipal Science and Technology Major Project by Anhui Initiative in Quantum Information Technologies. (Grant No. 2019SHZDZX01). We thank Dr. Xiaoxiao Wang for valuable advice during manuscript preparation. The experiments have been performed at the SX-ARPES endstation of the ADRESS beamline at the Swiss Light Source, Paul Scherrer Institute, Switzerland. Some preliminary data at the beginning of the project were taken at BL09U ARPES endstation of Shanghai Synchrotron Radiation Facility. 
\end{acknowledgments}

%\nocite{*}
%\bibliography{LAOKTO}% Produces the bibliography via BibTeX.

\begin{thebibliography}{65}%
\makeatletter
\providecommand \@ifxundefined [1]{%
 \@ifx{#1\undefined}
}%
\providecommand \@ifnum [1]{%
 \ifnum #1\expandafter \@firstoftwo
 \else \expandafter \@secondoftwo
 \fi
}%
\providecommand \@ifx [1]{%
 \ifx #1\expandafter \@firstoftwo
 \else \expandafter \@secondoftwo
 \fi
}%
\providecommand \natexlab [1]{#1}%
\providecommand \enquote  [1]{``#1''}%
\providecommand \bibnamefont  [1]{#1}%
\providecommand \bibfnamefont [1]{#1}%
\providecommand \citenamefont [1]{#1}%
\providecommand \href@noop [0]{\@secondoftwo}%
\providecommand \href [0]{\begingroup \@sanitize@url \@href}%
\providecommand \@href[1]{\@@startlink{#1}\@@href}%
\providecommand \@@href[1]{\endgroup#1\@@endlink}%
\providecommand \@sanitize@url [0]{\catcode `\\12\catcode `\$12\catcode
  `\&12\catcode `\#12\catcode `\^12\catcode `\_12\catcode `\%12\relax}%
\providecommand \@@startlink[1]{}%
\providecommand \@@endlink[0]{}%
\providecommand \url  [0]{\begingroup\@sanitize@url \@url }%
\providecommand \@url [1]{\endgroup\@href {#1}{\urlprefix }}%
\providecommand \urlprefix  [0]{URL }%
\providecommand \Eprint [0]{\href }%
\providecommand \doibase [0]{https://doi.org/}%
\providecommand \selectlanguage [0]{\@gobble}%
\providecommand \bibinfo  [0]{\@secondoftwo}%
\providecommand \bibfield  [0]{\@secondoftwo}%
\providecommand \translation [1]{[#1]}%
\providecommand \BibitemOpen [0]{}%
\providecommand \bibitemStop [0]{}%
\providecommand \bibitemNoStop [0]{.\EOS\space}%
\providecommand \EOS [0]{\spacefactor3000\relax}%
\providecommand \BibitemShut  [1]{\csname bibitem#1\endcsname}%
\let\auto@bib@innerbib\@empty
%</preamble>
\bibitem [{\citenamefont {Ohtomo}\ and\ \citenamefont {Hwang}(2004)}]{r1}%
  \BibitemOpen
  \bibfield  {author} {\bibinfo {author} {\bibfnamefont {A.}~\bibnamefont
  {Ohtomo}}\ and\ \bibinfo {author} {\bibfnamefont {H.~Y.}\ \bibnamefont
  {Hwang}},\ }\href {https://doi.org/10.1038/nature02308} {\bibfield  {journal}
  {\bibinfo  {journal} {Nature}\ }\textbf {\bibinfo {volume} {427}},\ \bibinfo
  {pages} {423} (\bibinfo {year} {2004})}\BibitemShut {NoStop}%
\bibitem [{\citenamefont {Reyren}\ \emph {et~al.}(2007)\citenamefont {Reyren},
  \citenamefont {Thiel}, \citenamefont {Caviglia}, \citenamefont {Kourkoutis},
  \citenamefont {Hammerl}, \citenamefont {Richter}, \citenamefont {Schneider},
  \citenamefont {Kopp}, \citenamefont {Ruetschi}, \citenamefont {Jaccard},
  \citenamefont {Gabay}, \citenamefont {Muller}, \citenamefont {Triscone},\
  and\ \citenamefont {Mannhart}}]{r2}%
  \BibitemOpen
  \bibfield  {author} {\bibinfo {author} {\bibfnamefont {N.}~\bibnamefont
  {Reyren}}, \bibinfo {author} {\bibfnamefont {S.}~\bibnamefont {Thiel}},
  \bibinfo {author} {\bibfnamefont {A.~D.}\ \bibnamefont {Caviglia}}, \bibinfo
  {author} {\bibfnamefont {L.~F.}\ \bibnamefont {Kourkoutis}}, \bibinfo
  {author} {\bibfnamefont {G.}~\bibnamefont {Hammerl}}, \bibinfo {author}
  {\bibfnamefont {C.}~\bibnamefont {Richter}}, \bibinfo {author} {\bibfnamefont
  {C.~W.}\ \bibnamefont {Schneider}}, \bibinfo {author} {\bibfnamefont
  {T.}~\bibnamefont {Kopp}}, \bibinfo {author} {\bibfnamefont {A.~S.}\
  \bibnamefont {Ruetschi}}, \bibinfo {author} {\bibfnamefont {D.}~\bibnamefont
  {Jaccard}}, \bibinfo {author} {\bibfnamefont {M.}~\bibnamefont {Gabay}},
  \bibinfo {author} {\bibfnamefont {D.~A.}\ \bibnamefont {Muller}}, \bibinfo
  {author} {\bibfnamefont {J.~M.}\ \bibnamefont {Triscone}},\ and\ \bibinfo
  {author} {\bibfnamefont {J.}~\bibnamefont {Mannhart}},\ }\href
  {https://doi.org/10.1126/science.1146006} {\bibfield  {journal} {\bibinfo
  {journal} {Science}\ }\textbf {\bibinfo {volume} {317}},\ \bibinfo {pages}
  {1196} (\bibinfo {year} {2007})}\BibitemShut {NoStop}%
\bibitem [{\citenamefont {Gastiasoro}\ \emph {et~al.}(2020)\citenamefont
  {Gastiasoro}, \citenamefont {Ruhman},\ and\ \citenamefont {Fernandes}}]{r3}%
  \BibitemOpen
  \bibfield  {author} {\bibinfo {author} {\bibfnamefont {M.~N.}\ \bibnamefont
  {Gastiasoro}}, \bibinfo {author} {\bibfnamefont {J.}~\bibnamefont {Ruhman}},\
  and\ \bibinfo {author} {\bibfnamefont {R.~M.}\ \bibnamefont {Fernandes}},\
  }\href {https://doi.org/10.1016/j.aop.2020.168107} {\bibfield  {journal}
  {\bibinfo  {journal} {Annals of Physics}\ }\textbf {\bibinfo {volume}
  {417}},\ \bibinfo {pages} {168107} (\bibinfo {year} {2020})}\BibitemShut
  {NoStop}%
\bibitem [{\citenamefont {Wölfle}\ and\ \citenamefont {Balatsky}(2018)}]{r4}%
  \BibitemOpen
  \bibfield  {author} {\bibinfo {author} {\bibfnamefont {P.}~\bibnamefont
  {Wölfle}}\ and\ \bibinfo {author} {\bibfnamefont {A.~V.}\ \bibnamefont
  {Balatsky}},\ }\href {https://doi.org/10.1103/PhysRevB.98.104505} {\bibfield
  {journal} {\bibinfo  {journal} {Phys. Rev. B}\ }\textbf {\bibinfo {volume}
  {98}},\ \bibinfo {pages} {104505} (\bibinfo {year} {2018})}\BibitemShut
  {NoStop}%
\bibitem [{\citenamefont {Enderlein}\ \emph {et~al.}(2020)\citenamefont
  {Enderlein}, \citenamefont {de~Oliveira}, \citenamefont {Tompsett},
  \citenamefont {Saitovitch}, \citenamefont {Saxena}, \citenamefont
  {Lonzarich},\ and\ \citenamefont {Rowley}}]{r5}%
  \BibitemOpen
  \bibfield  {author} {\bibinfo {author} {\bibfnamefont {C.}~\bibnamefont
  {Enderlein}}, \bibinfo {author} {\bibfnamefont {J.~F.}\ \bibnamefont
  {de~Oliveira}}, \bibinfo {author} {\bibfnamefont {D.~A.}\ \bibnamefont
  {Tompsett}}, \bibinfo {author} {\bibfnamefont {E.~B.}\ \bibnamefont
  {Saitovitch}}, \bibinfo {author} {\bibfnamefont {S.~S.}\ \bibnamefont
  {Saxena}}, \bibinfo {author} {\bibfnamefont {G.~G.}\ \bibnamefont
  {Lonzarich}},\ and\ \bibinfo {author} {\bibfnamefont {S.~E.}\ \bibnamefont
  {Rowley}},\ }\href {https://doi.org/10.1038/s41467-020-18438-0} {\bibfield
  {journal} {\bibinfo  {journal} {Nat. Commun.}\ }\textbf {\bibinfo {volume}
  {11}},\ \bibinfo {pages} {4852} (\bibinfo {year} {2020})}\BibitemShut
  {NoStop}%
\bibitem [{\citenamefont {van~der Marel}\ \emph {et~al.}(2019)\citenamefont
  {van~der Marel}, \citenamefont {Barantani},\ and\ \citenamefont
  {Rischau}}]{r6}%
  \BibitemOpen
  \bibfield  {author} {\bibinfo {author} {\bibfnamefont {D.}~\bibnamefont
  {van~der Marel}}, \bibinfo {author} {\bibfnamefont {F.}~\bibnamefont
  {Barantani}},\ and\ \bibinfo {author} {\bibfnamefont {C.~W.}\ \bibnamefont
  {Rischau}},\ }\href {https://doi.org/10.1103/PhysRevResearch.1.013003}
  {\bibfield  {journal} {\bibinfo  {journal} {Phys. Rev. Research}\ }\textbf
  {\bibinfo {volume} {1}},\ \bibinfo {pages} {013003} (\bibinfo {year}
  {2019})}\BibitemShut {NoStop}%
\bibitem [{\citenamefont {Cancellieri}\ \emph {et~al.}(2016)\citenamefont
  {Cancellieri}, \citenamefont {Mishchenko}, \citenamefont {Aschauer},
  \citenamefont {Filippetti}, \citenamefont {Faber}, \citenamefont {Barisic},
  \citenamefont {Rogalev}, \citenamefont {Schmitt}, \citenamefont {Nagaosa},\
  and\ \citenamefont {Strocov}}]{r7}%
  \BibitemOpen
  \bibfield  {author} {\bibinfo {author} {\bibfnamefont {C.}~\bibnamefont
  {Cancellieri}}, \bibinfo {author} {\bibfnamefont {A.~S.}\ \bibnamefont
  {Mishchenko}}, \bibinfo {author} {\bibfnamefont {U.}~\bibnamefont
  {Aschauer}}, \bibinfo {author} {\bibfnamefont {A.}~\bibnamefont
  {Filippetti}}, \bibinfo {author} {\bibfnamefont {C.}~\bibnamefont {Faber}},
  \bibinfo {author} {\bibfnamefont {O.~S.}\ \bibnamefont {Barisic}}, \bibinfo
  {author} {\bibfnamefont {V.~A.}\ \bibnamefont {Rogalev}}, \bibinfo {author}
  {\bibfnamefont {T.}~\bibnamefont {Schmitt}}, \bibinfo {author} {\bibfnamefont
  {N.}~\bibnamefont {Nagaosa}},\ and\ \bibinfo {author} {\bibfnamefont {V.~N.}\
  \bibnamefont {Strocov}},\ }\href {https://doi.org/10.1038/ncomms10386}
  {\bibfield  {journal} {\bibinfo  {journal} {Nat. Commun.}\ }\textbf {\bibinfo
  {volume} {7}},\ \bibinfo {pages} {10386} (\bibinfo {year}
  {2016})}\BibitemShut {NoStop}%
\bibitem [{\citenamefont {Wang}\ \emph {et~al.}(2016)\citenamefont {Wang},
  \citenamefont {McKeown~Walker}, \citenamefont {Tamai}, \citenamefont {Wang},
  \citenamefont {Ristic}, \citenamefont {Bruno}, \citenamefont {de~la Torre},
  \citenamefont {Ricco}, \citenamefont {Plumb}, \citenamefont {Shi},
  \citenamefont {Hlawenka}, \citenamefont {Sanchez-Barriga}, \citenamefont
  {Varykhalov}, \citenamefont {Kim}, \citenamefont {Hoesch}, \citenamefont
  {King}, \citenamefont {Meevasana}, \citenamefont {Diebold}, \citenamefont
  {Mesot}, \citenamefont {Moritz}, \citenamefont {Devereaux}, \citenamefont
  {Radovic},\ and\ \citenamefont {Baumberger}}]{r8}%
  \BibitemOpen
  \bibfield  {author} {\bibinfo {author} {\bibfnamefont {Z.}~\bibnamefont
  {Wang}}, \bibinfo {author} {\bibfnamefont {S.}~\bibnamefont
  {McKeown~Walker}}, \bibinfo {author} {\bibfnamefont {A.}~\bibnamefont
  {Tamai}}, \bibinfo {author} {\bibfnamefont {Y.}~\bibnamefont {Wang}},
  \bibinfo {author} {\bibfnamefont {Z.}~\bibnamefont {Ristic}}, \bibinfo
  {author} {\bibfnamefont {F.~Y.}\ \bibnamefont {Bruno}}, \bibinfo {author}
  {\bibfnamefont {A.}~\bibnamefont {de~la Torre}}, \bibinfo {author}
  {\bibfnamefont {S.}~\bibnamefont {Ricco}}, \bibinfo {author} {\bibfnamefont
  {N.~C.}\ \bibnamefont {Plumb}}, \bibinfo {author} {\bibfnamefont
  {M.}~\bibnamefont {Shi}}, \bibinfo {author} {\bibfnamefont {P.}~\bibnamefont
  {Hlawenka}}, \bibinfo {author} {\bibfnamefont {J.}~\bibnamefont
  {Sanchez-Barriga}}, \bibinfo {author} {\bibfnamefont {A.}~\bibnamefont
  {Varykhalov}}, \bibinfo {author} {\bibfnamefont {T.~K.}\ \bibnamefont {Kim}},
  \bibinfo {author} {\bibfnamefont {M.}~\bibnamefont {Hoesch}}, \bibinfo
  {author} {\bibfnamefont {P.~D.}\ \bibnamefont {King}}, \bibinfo {author}
  {\bibfnamefont {W.}~\bibnamefont {Meevasana}}, \bibinfo {author}
  {\bibfnamefont {U.}~\bibnamefont {Diebold}}, \bibinfo {author} {\bibfnamefont
  {J.}~\bibnamefont {Mesot}}, \bibinfo {author} {\bibfnamefont
  {B.}~\bibnamefont {Moritz}}, \bibinfo {author} {\bibfnamefont {T.~P.}\
  \bibnamefont {Devereaux}}, \bibinfo {author} {\bibfnamefont {M.}~\bibnamefont
  {Radovic}},\ and\ \bibinfo {author} {\bibfnamefont {F.}~\bibnamefont
  {Baumberger}},\ }\href {https://doi.org/10.1038/nmat4623} {\bibfield
  {journal} {\bibinfo  {journal} {Nat. Mater.}\ }\textbf {\bibinfo {volume}
  {15}},\ \bibinfo {pages} {835} (\bibinfo {year} {2016})}\BibitemShut
  {NoStop}%
\bibitem [{\citenamefont {Chen}\ \emph {et~al.}(2015)\citenamefont {Chen},
  \citenamefont {Avila}, \citenamefont {Frantzeskakis}, \citenamefont {Levy},\
  and\ \citenamefont {Asensio}}]{r9}%
  \BibitemOpen
  \bibfield  {author} {\bibinfo {author} {\bibfnamefont {C.}~\bibnamefont
  {Chen}}, \bibinfo {author} {\bibfnamefont {J.}~\bibnamefont {Avila}},
  \bibinfo {author} {\bibfnamefont {E.}~\bibnamefont {Frantzeskakis}}, \bibinfo
  {author} {\bibfnamefont {A.}~\bibnamefont {Levy}},\ and\ \bibinfo {author}
  {\bibfnamefont {M.~C.}\ \bibnamefont {Asensio}},\ }\href
  {https://doi.org/10.1038/ncomms9585} {\bibfield  {journal} {\bibinfo
  {journal} {Nat. Commun.}\ }\textbf {\bibinfo {volume} {6}},\ \bibinfo {pages}
  {8585} (\bibinfo {year} {2015})}\BibitemShut {NoStop}%
\bibitem [{\citenamefont {Meevasana}\ \emph {et~al.}(2010)\citenamefont
  {Meevasana}, \citenamefont {Zhou}, \citenamefont {Moritz}, \citenamefont
  {Chen}, \citenamefont {He}, \citenamefont {Fujimori}, \citenamefont {Lu},
  \citenamefont {Mo}, \citenamefont {Moore}, \citenamefont {Baumberger},
  \citenamefont {Devereaux}, \citenamefont {van~der Marel}, \citenamefont
  {Nagaosa}, \citenamefont {Zaanen},\ and\ \citenamefont {Shen}}]{r10}%
  \BibitemOpen
  \bibfield  {author} {\bibinfo {author} {\bibfnamefont {W.}~\bibnamefont
  {Meevasana}}, \bibinfo {author} {\bibfnamefont {X.~J.}\ \bibnamefont {Zhou}},
  \bibinfo {author} {\bibfnamefont {B.}~\bibnamefont {Moritz}}, \bibinfo
  {author} {\bibfnamefont {C.~C.}\ \bibnamefont {Chen}}, \bibinfo {author}
  {\bibfnamefont {R.~H.}\ \bibnamefont {He}}, \bibinfo {author} {\bibfnamefont
  {S.~I.}\ \bibnamefont {Fujimori}}, \bibinfo {author} {\bibfnamefont {D.~H.}\
  \bibnamefont {Lu}}, \bibinfo {author} {\bibfnamefont {S.~K.}\ \bibnamefont
  {Mo}}, \bibinfo {author} {\bibfnamefont {R.~G.}\ \bibnamefont {Moore}},
  \bibinfo {author} {\bibfnamefont {F.}~\bibnamefont {Baumberger}}, \bibinfo
  {author} {\bibfnamefont {T.~P.}\ \bibnamefont {Devereaux}}, \bibinfo {author}
  {\bibfnamefont {D.}~\bibnamefont {van~der Marel}}, \bibinfo {author}
  {\bibfnamefont {N.}~\bibnamefont {Nagaosa}}, \bibinfo {author} {\bibfnamefont
  {J.}~\bibnamefont {Zaanen}},\ and\ \bibinfo {author} {\bibfnamefont {Z.~X.}\
  \bibnamefont {Shen}},\ }\href {https://doi.org/10.1088/1367-2630/12/2/023004}
  {\bibfield  {journal} {\bibinfo  {journal} {New J. Phys.}\ }\textbf {\bibinfo
  {volume} {12}},\ \bibinfo {pages} {023004} (\bibinfo {year}
  {2010})}\BibitemShut {NoStop}%
\bibitem [{\citenamefont {Geondzhian}\ \emph {et~al.}(2020)\citenamefont
  {Geondzhian}, \citenamefont {Sambri}, \citenamefont {De~Luca}, \citenamefont
  {Di~Capua}, \citenamefont {Di~Gennaro}, \citenamefont {Betto}, \citenamefont
  {Rossi}, \citenamefont {Peng}, \citenamefont {Fumagalli}, \citenamefont
  {Brookes}, \citenamefont {Braicovich}, \citenamefont {Gilmore}, \citenamefont
  {Ghiringhelli},\ and\ \citenamefont {Salluzzo}}]{r11}%
  \BibitemOpen
  \bibfield  {author} {\bibinfo {author} {\bibfnamefont {A.}~\bibnamefont
  {Geondzhian}}, \bibinfo {author} {\bibfnamefont {A.}~\bibnamefont {Sambri}},
  \bibinfo {author} {\bibfnamefont {G.~M.}\ \bibnamefont {De~Luca}}, \bibinfo
  {author} {\bibfnamefont {R.}~\bibnamefont {Di~Capua}}, \bibinfo {author}
  {\bibfnamefont {E.}~\bibnamefont {Di~Gennaro}}, \bibinfo {author}
  {\bibfnamefont {D.}~\bibnamefont {Betto}}, \bibinfo {author} {\bibfnamefont
  {M.}~\bibnamefont {Rossi}}, \bibinfo {author} {\bibfnamefont {Y.~Y.}\
  \bibnamefont {Peng}}, \bibinfo {author} {\bibfnamefont {R.}~\bibnamefont
  {Fumagalli}}, \bibinfo {author} {\bibfnamefont {N.~B.}\ \bibnamefont
  {Brookes}}, \bibinfo {author} {\bibfnamefont {L.}~\bibnamefont {Braicovich}},
  \bibinfo {author} {\bibfnamefont {K.}~\bibnamefont {Gilmore}}, \bibinfo
  {author} {\bibfnamefont {G.}~\bibnamefont {Ghiringhelli}},\ and\ \bibinfo
  {author} {\bibfnamefont {M.}~\bibnamefont {Salluzzo}},\ }\href
  {https://doi.org/10.1103/PhysRevLett.125.126401} {\bibfield  {journal}
  {\bibinfo  {journal} {Phys. Rev. Lett.}\ }\textbf {\bibinfo {volume} {125}},\
  \bibinfo {pages} {126401} (\bibinfo {year} {2020})}\BibitemShut {NoStop}%
\bibitem [{\citenamefont {Gor'kov}(2016)}]{r12}%
  \BibitemOpen
  \bibfield  {author} {\bibinfo {author} {\bibfnamefont {L.~P.}\ \bibnamefont
  {Gor'kov}},\ }\href {https://doi.org/10.1073/pnas.1604145113} {\bibfield
  {journal} {\bibinfo  {journal} {Proc. Natl. Acad. Sci. U.S.A.}\ }\textbf
  {\bibinfo {volume} {113}},\ \bibinfo {pages} {4646} (\bibinfo {year}
  {2016})}\BibitemShut {NoStop}%
\bibitem [{\citenamefont {Klimin}\ \emph {et~al.}(2014)\citenamefont {Klimin},
  \citenamefont {Tempere}, \citenamefont {Devreese},\ and\ \citenamefont
  {van~der Marel}}]{r13}%
  \BibitemOpen
  \bibfield  {author} {\bibinfo {author} {\bibfnamefont {S.~N.}\ \bibnamefont
  {Klimin}}, \bibinfo {author} {\bibfnamefont {J.}~\bibnamefont {Tempere}},
  \bibinfo {author} {\bibfnamefont {J.~T.}\ \bibnamefont {Devreese}},\ and\
  \bibinfo {author} {\bibfnamefont {D.}~\bibnamefont {van~der Marel}},\ }\href
  {https://doi.org/10.1103/PhysRevB.89.184514} {\bibfield  {journal} {\bibinfo
  {journal} {Phys. Rev. B}\ }\textbf {\bibinfo {volume} {89}},\ \bibinfo
  {pages} {184514} (\bibinfo {year} {2014})}\BibitemShut {NoStop}%
\bibitem [{\citenamefont {Liu}\ \emph {et~al.}(2021{\natexlab{a}})\citenamefont
  {Liu}, \citenamefont {Yan}, \citenamefont {Jin}, \citenamefont {Ma},
  \citenamefont {Hsiao}, \citenamefont {Lin}, \citenamefont {Bretz-Sullivan},
  \citenamefont {Zhou}, \citenamefont {Pearson}, \citenamefont {Fisher},
  \citenamefont {Jiang}, \citenamefont {Han}, \citenamefont {Zuo},
  \citenamefont {Wen}, \citenamefont {Fong}, \citenamefont {Sun}, \citenamefont
  {Zhou},\ and\ \citenamefont {Bhattacharya}}]{r14}%
  \BibitemOpen
  \bibfield  {author} {\bibinfo {author} {\bibfnamefont {C.}~\bibnamefont
  {Liu}}, \bibinfo {author} {\bibfnamefont {X.}~\bibnamefont {Yan}}, \bibinfo
  {author} {\bibfnamefont {D.}~\bibnamefont {Jin}}, \bibinfo {author}
  {\bibfnamefont {Y.}~\bibnamefont {Ma}}, \bibinfo {author} {\bibfnamefont
  {H.~W.}\ \bibnamefont {Hsiao}}, \bibinfo {author} {\bibfnamefont
  {Y.}~\bibnamefont {Lin}}, \bibinfo {author} {\bibfnamefont {T.~M.}\
  \bibnamefont {Bretz-Sullivan}}, \bibinfo {author} {\bibfnamefont
  {X.}~\bibnamefont {Zhou}}, \bibinfo {author} {\bibfnamefont {J.}~\bibnamefont
  {Pearson}}, \bibinfo {author} {\bibfnamefont {B.}~\bibnamefont {Fisher}},
  \bibinfo {author} {\bibfnamefont {J.~S.}\ \bibnamefont {Jiang}}, \bibinfo
  {author} {\bibfnamefont {W.}~\bibnamefont {Han}}, \bibinfo {author}
  {\bibfnamefont {J.~M.}\ \bibnamefont {Zuo}}, \bibinfo {author} {\bibfnamefont
  {J.}~\bibnamefont {Wen}}, \bibinfo {author} {\bibfnamefont {D.~D.}\
  \bibnamefont {Fong}}, \bibinfo {author} {\bibfnamefont {J.}~\bibnamefont
  {Sun}}, \bibinfo {author} {\bibfnamefont {H.}~\bibnamefont {Zhou}},\ and\
  \bibinfo {author} {\bibfnamefont {A.}~\bibnamefont {Bhattacharya}},\ }\href
  {https://doi.org/10.1126/science.aba5511} {\bibfield  {journal} {\bibinfo
  {journal} {Science}\ }\textbf {\bibinfo {volume} {371}},\ \bibinfo {pages}
  {716} (\bibinfo {year} {2021}{\natexlab{a}})}\BibitemShut {NoStop}%
\bibitem [{\citenamefont {Chen}\ \emph
  {et~al.}(2021{\natexlab{a}})\citenamefont {Chen}, \citenamefont {Liu},
  \citenamefont {Zhang}, \citenamefont {Liu}, \citenamefont {Tian},
  \citenamefont {Sun}, \citenamefont {Zhang}, \citenamefont {Zhou},
  \citenamefont {Sun},\ and\ \citenamefont {Xie}}]{r15}%
  \BibitemOpen
  \bibfield  {author} {\bibinfo {author} {\bibfnamefont {Z.}~\bibnamefont
  {Chen}}, \bibinfo {author} {\bibfnamefont {Y.}~\bibnamefont {Liu}}, \bibinfo
  {author} {\bibfnamefont {H.}~\bibnamefont {Zhang}}, \bibinfo {author}
  {\bibfnamefont {Z.}~\bibnamefont {Liu}}, \bibinfo {author} {\bibfnamefont
  {H.}~\bibnamefont {Tian}}, \bibinfo {author} {\bibfnamefont {Y.}~\bibnamefont
  {Sun}}, \bibinfo {author} {\bibfnamefont {M.}~\bibnamefont {Zhang}}, \bibinfo
  {author} {\bibfnamefont {Y.}~\bibnamefont {Zhou}}, \bibinfo {author}
  {\bibfnamefont {J.}~\bibnamefont {Sun}},\ and\ \bibinfo {author}
  {\bibfnamefont {Y.}~\bibnamefont {Xie}},\ }\href
  {https://doi.org/10.1126/science.abb3848} {\bibfield  {journal} {\bibinfo
  {journal} {Science}\ }\textbf {\bibinfo {volume} {372}},\ \bibinfo {pages}
  {721} (\bibinfo {year} {2021}{\natexlab{a}})}\BibitemShut {NoStop}%
\bibitem [{\citenamefont {Qiao}\ \emph {et~al.}(2021)\citenamefont {Qiao},
  \citenamefont {Ma}, \citenamefont {Yan}, \citenamefont {Xing}, \citenamefont
  {Yao}, \citenamefont {Cai}, \citenamefont {Li}, \citenamefont {Xiong},
  \citenamefont {Xie}, \citenamefont {Lin},\ and\ \citenamefont {Han}}]{r16}%
  \BibitemOpen
  \bibfield  {author} {\bibinfo {author} {\bibfnamefont {W.}~\bibnamefont
  {Qiao}}, \bibinfo {author} {\bibfnamefont {Y.}~\bibnamefont {Ma}}, \bibinfo
  {author} {\bibfnamefont {J.}~\bibnamefont {Yan}}, \bibinfo {author}
  {\bibfnamefont {W.}~\bibnamefont {Xing}}, \bibinfo {author} {\bibfnamefont
  {Y.}~\bibnamefont {Yao}}, \bibinfo {author} {\bibfnamefont {R.}~\bibnamefont
  {Cai}}, \bibinfo {author} {\bibfnamefont {B.}~\bibnamefont {Li}}, \bibinfo
  {author} {\bibfnamefont {R.}~\bibnamefont {Xiong}}, \bibinfo {author}
  {\bibfnamefont {X.~C.}\ \bibnamefont {Xie}}, \bibinfo {author} {\bibfnamefont
  {X.}~\bibnamefont {Lin}},\ and\ \bibinfo {author} {\bibfnamefont
  {W.}~\bibnamefont {Han}},\ }\href
  {https://doi.org/10.1103/PhysRevB.104.184505} {\bibfield  {journal} {\bibinfo
   {journal} {Phys. Rev. B}\ }\textbf {\bibinfo {volume} {104}},\ \bibinfo
  {pages} {184505} (\bibinfo {year} {2021})}\BibitemShut {NoStop}%
\bibitem [{\citenamefont {Rubi}\ \emph {et~al.}(2021)\citenamefont {Rubi},
  \citenamefont {Zeng}, \citenamefont {Bangma}, \citenamefont {Goiran},
  \citenamefont {Ariando}, \citenamefont {Escoffier},\ and\ \citenamefont
  {Zeitler}}]{r17}%
  \BibitemOpen
  \bibfield  {author} {\bibinfo {author} {\bibfnamefont {K.}~\bibnamefont
  {Rubi}}, \bibinfo {author} {\bibfnamefont {S.}~\bibnamefont {Zeng}}, \bibinfo
  {author} {\bibfnamefont {F.}~\bibnamefont {Bangma}}, \bibinfo {author}
  {\bibfnamefont {M.}~\bibnamefont {Goiran}}, \bibinfo {author} {\bibfnamefont
  {A.}~\bibnamefont {Ariando}}, \bibinfo {author} {\bibfnamefont
  {W.}~\bibnamefont {Escoffier}},\ and\ \bibinfo {author} {\bibfnamefont
  {U.}~\bibnamefont {Zeitler}},\ }\href
  {https://doi.org/10.1103/PhysRevResearch.3.033234} {\bibfield  {journal}
  {\bibinfo  {journal} {Phys. Rev. Research}\ }\textbf {\bibinfo {volume}
  {3}},\ \bibinfo {pages} {033234} (\bibinfo {year} {2021})}\BibitemShut
  {NoStop}%
\bibitem [{\citenamefont {Zapf}\ \emph {et~al.}(2022)\citenamefont {Zapf},
  \citenamefont {Schmitt}, \citenamefont {Gabel}, \citenamefont {Scheiderer},
  \citenamefont {Stübinger}, \citenamefont {Leikert}, \citenamefont
  {Sangiovanni}, \citenamefont {Dudy}, \citenamefont {Chernov}, \citenamefont
  {Babenkov}, \citenamefont {Vasilyev}, \citenamefont {Fedchenko},
  \citenamefont {Medjanik}, \citenamefont {Matveyev}, \citenamefont
  {Gloskowski}, \citenamefont {Schlueter}, \citenamefont {Lee}, \citenamefont
  {Elmers}, \citenamefont {Schönhense}, \citenamefont {Sing},\ and\
  \citenamefont {Claessen}}]{r18}%
  \BibitemOpen
  \bibfield  {author} {\bibinfo {author} {\bibfnamefont {M.}~\bibnamefont
  {Zapf}}, \bibinfo {author} {\bibfnamefont {M.}~\bibnamefont {Schmitt}},
  \bibinfo {author} {\bibfnamefont {J.}~\bibnamefont {Gabel}}, \bibinfo
  {author} {\bibfnamefont {P.}~\bibnamefont {Scheiderer}}, \bibinfo {author}
  {\bibfnamefont {M.}~\bibnamefont {Stübinger}}, \bibinfo {author}
  {\bibfnamefont {B.}~\bibnamefont {Leikert}}, \bibinfo {author} {\bibfnamefont
  {G.}~\bibnamefont {Sangiovanni}}, \bibinfo {author} {\bibfnamefont
  {L.}~\bibnamefont {Dudy}}, \bibinfo {author} {\bibfnamefont {S.}~\bibnamefont
  {Chernov}}, \bibinfo {author} {\bibfnamefont {S.}~\bibnamefont {Babenkov}},
  \bibinfo {author} {\bibfnamefont {D.}~\bibnamefont {Vasilyev}}, \bibinfo
  {author} {\bibfnamefont {O.}~\bibnamefont {Fedchenko}}, \bibinfo {author}
  {\bibfnamefont {K.}~\bibnamefont {Medjanik}}, \bibinfo {author}
  {\bibfnamefont {Y.}~\bibnamefont {Matveyev}}, \bibinfo {author}
  {\bibfnamefont {A.}~\bibnamefont {Gloskowski}}, \bibinfo {author}
  {\bibfnamefont {C.}~\bibnamefont {Schlueter}}, \bibinfo {author}
  {\bibfnamefont {T.~L.}\ \bibnamefont {Lee}}, \bibinfo {author} {\bibfnamefont
  {H.~J.}\ \bibnamefont {Elmers}}, \bibinfo {author} {\bibfnamefont
  {G.}~\bibnamefont {Schönhense}}, \bibinfo {author} {\bibfnamefont
  {M.}~\bibnamefont {Sing}},\ and\ \bibinfo {author} {\bibfnamefont
  {R.}~\bibnamefont {Claessen}},\ }\href
  {https://doi.org/10.1103/PhysRevB.106.125137} {\bibfield  {journal} {\bibinfo
   {journal} {Phys. Rev. B}\ }\textbf {\bibinfo {volume} {106}},\ \bibinfo
  {pages} {125137} (\bibinfo {year} {2022})}\BibitemShut {NoStop}%
\bibitem [{\citenamefont {Chen}\ \emph
  {et~al.}(2021{\natexlab{b}})\citenamefont {Chen}, \citenamefont {Liu},
  \citenamefont {Sun}, \citenamefont {Chen}, \citenamefont {Liu}, \citenamefont
  {Zhang}, \citenamefont {Li}, \citenamefont {Zhang}, \citenamefont {Hong},
  \citenamefont {Ren}, \citenamefont {Zhang}, \citenamefont {Tian},
  \citenamefont {Zhou}, \citenamefont {Sun},\ and\ \citenamefont {Xie}}]{r19}%
  \BibitemOpen
  \bibfield  {author} {\bibinfo {author} {\bibfnamefont {Z.}~\bibnamefont
  {Chen}}, \bibinfo {author} {\bibfnamefont {Z.}~\bibnamefont {Liu}}, \bibinfo
  {author} {\bibfnamefont {Y.}~\bibnamefont {Sun}}, \bibinfo {author}
  {\bibfnamefont {X.}~\bibnamefont {Chen}}, \bibinfo {author} {\bibfnamefont
  {Y.}~\bibnamefont {Liu}}, \bibinfo {author} {\bibfnamefont {H.}~\bibnamefont
  {Zhang}}, \bibinfo {author} {\bibfnamefont {H.}~\bibnamefont {Li}}, \bibinfo
  {author} {\bibfnamefont {M.}~\bibnamefont {Zhang}}, \bibinfo {author}
  {\bibfnamefont {S.}~\bibnamefont {Hong}}, \bibinfo {author} {\bibfnamefont
  {T.}~\bibnamefont {Ren}}, \bibinfo {author} {\bibfnamefont {C.}~\bibnamefont
  {Zhang}}, \bibinfo {author} {\bibfnamefont {H.}~\bibnamefont {Tian}},
  \bibinfo {author} {\bibfnamefont {Y.}~\bibnamefont {Zhou}}, \bibinfo {author}
  {\bibfnamefont {J.}~\bibnamefont {Sun}},\ and\ \bibinfo {author}
  {\bibfnamefont {Y.}~\bibnamefont {Xie}},\ }\href
  {https://doi.org/10.1103/PhysRevLett.126.026802} {\bibfield  {journal}
  {\bibinfo  {journal} {Phys. Rev. Lett.}\ }\textbf {\bibinfo {volume} {126}},\
  \bibinfo {pages} {026802} (\bibinfo {year} {2021}{\natexlab{b}})}\BibitemShut
  {NoStop}%
\bibitem [{\citenamefont {Mallik}\ \emph {et~al.}(2022)\citenamefont {Mallik},
  \citenamefont {Menard}, \citenamefont {Saiz}, \citenamefont {Witt},
  \citenamefont {Lesueur}, \citenamefont {Gloter}, \citenamefont {Benfatto},
  \citenamefont {Bibes},\ and\ \citenamefont {Bergeal}}]{r20}%
  \BibitemOpen
  \bibfield  {author} {\bibinfo {author} {\bibfnamefont {S.}~\bibnamefont
  {Mallik}}, \bibinfo {author} {\bibfnamefont {G.~C.}\ \bibnamefont {Menard}},
  \bibinfo {author} {\bibfnamefont {G.}~\bibnamefont {Saiz}}, \bibinfo {author}
  {\bibfnamefont {H.}~\bibnamefont {Witt}}, \bibinfo {author} {\bibfnamefont
  {J.}~\bibnamefont {Lesueur}}, \bibinfo {author} {\bibfnamefont
  {A.}~\bibnamefont {Gloter}}, \bibinfo {author} {\bibfnamefont
  {L.}~\bibnamefont {Benfatto}}, \bibinfo {author} {\bibfnamefont
  {M.}~\bibnamefont {Bibes}},\ and\ \bibinfo {author} {\bibfnamefont
  {N.}~\bibnamefont {Bergeal}},\ }\href
  {https://doi.org/10.1038/s41467-022-32242-y} {\bibfield  {journal} {\bibinfo
  {journal} {Nat. Commun.}\ }\textbf {\bibinfo {volume} {13}},\ \bibinfo
  {pages} {4625} (\bibinfo {year} {2022})}\BibitemShut {NoStop}%
\bibitem [{\citenamefont {Liu}\ \emph {et~al.}(2022)\citenamefont {Liu},
  \citenamefont {Zhou}, \citenamefont {Hong}, \citenamefont {Fisher},
  \citenamefont {Zheng}, \citenamefont {Pearson}, \citenamefont {Jin},
  \citenamefont {Norman},\ and\ \citenamefont {Bhattacharya}}]{r21}%
  \BibitemOpen
  \bibfield  {author} {\bibinfo {author} {\bibfnamefont {C.}~\bibnamefont
  {Liu}}, \bibinfo {author} {\bibfnamefont {X.}~\bibnamefont {Zhou}}, \bibinfo
  {author} {\bibfnamefont {D.}~\bibnamefont {Hong}}, \bibinfo {author}
  {\bibfnamefont {B.}~\bibnamefont {Fisher}}, \bibinfo {author} {\bibfnamefont
  {H.}~\bibnamefont {Zheng}}, \bibinfo {author} {\bibfnamefont
  {J.}~\bibnamefont {Pearson}}, \bibinfo {author} {\bibfnamefont
  {D.}~\bibnamefont {Jin}}, \bibinfo {author} {\bibfnamefont {M.~R.}\
  \bibnamefont {Norman}},\ and\ \bibinfo {author} {\bibfnamefont
  {A.}~\bibnamefont {Bhattacharya}}} (\bibinfo {year} {2022}),\ \bibinfo {note}
  {preprint at https://arxiv.org/abs/2203.05867}\BibitemShut {NoStop}%
\bibitem [{\citenamefont {Popovic}\ \emph {et~al.}(2008)\citenamefont
  {Popovic}, \citenamefont {Satpathy},\ and\ \citenamefont {Martin}}]{r22}%
  \BibitemOpen
  \bibfield  {author} {\bibinfo {author} {\bibfnamefont {Z.~S.}\ \bibnamefont
  {Popovic}}, \bibinfo {author} {\bibfnamefont {S.}~\bibnamefont {Satpathy}},\
  and\ \bibinfo {author} {\bibfnamefont {R.~M.}\ \bibnamefont {Martin}},\
  }\href {https://doi.org/10.1103/PhysRevLett.101.256801} {\bibfield  {journal}
  {\bibinfo  {journal} {Phys. Rev. Lett.}\ }\textbf {\bibinfo {volume} {101}},\
  \bibinfo {pages} {256801} (\bibinfo {year} {2008})}\BibitemShut {NoStop}%
\bibitem [{\citenamefont {Sing}\ \emph {et~al.}(2009)\citenamefont {Sing},
  \citenamefont {Berner}, \citenamefont {Goss}, \citenamefont {Muller},
  \citenamefont {Ruff}, \citenamefont {Wetscherek}, \citenamefont {Thiel},
  \citenamefont {Mannhart}, \citenamefont {Pauli}, \citenamefont {Schneider},
  \citenamefont {Willmott}, \citenamefont {Gorgoi}, \citenamefont {Schafers},\
  and\ \citenamefont {Claessen}}]{r23}%
  \BibitemOpen
  \bibfield  {author} {\bibinfo {author} {\bibfnamefont {M.}~\bibnamefont
  {Sing}}, \bibinfo {author} {\bibfnamefont {G.}~\bibnamefont {Berner}},
  \bibinfo {author} {\bibfnamefont {K.}~\bibnamefont {Goss}}, \bibinfo {author}
  {\bibfnamefont {A.}~\bibnamefont {Muller}}, \bibinfo {author} {\bibfnamefont
  {A.}~\bibnamefont {Ruff}}, \bibinfo {author} {\bibfnamefont {A.}~\bibnamefont
  {Wetscherek}}, \bibinfo {author} {\bibfnamefont {S.}~\bibnamefont {Thiel}},
  \bibinfo {author} {\bibfnamefont {J.}~\bibnamefont {Mannhart}}, \bibinfo
  {author} {\bibfnamefont {S.~A.}\ \bibnamefont {Pauli}}, \bibinfo {author}
  {\bibfnamefont {C.~W.}\ \bibnamefont {Schneider}}, \bibinfo {author}
  {\bibfnamefont {P.~R.}\ \bibnamefont {Willmott}}, \bibinfo {author}
  {\bibfnamefont {M.}~\bibnamefont {Gorgoi}}, \bibinfo {author} {\bibfnamefont
  {F.}~\bibnamefont {Schafers}},\ and\ \bibinfo {author} {\bibfnamefont
  {R.}~\bibnamefont {Claessen}},\ }\href
  {https://doi.org/10.1103/PhysRevLett.102.176805} {\bibfield  {journal}
  {\bibinfo  {journal} {Phys. Rev. Lett.}\ }\textbf {\bibinfo {volume} {102}},\
  \bibinfo {pages} {176805} (\bibinfo {year} {2009})}\BibitemShut {NoStop}%
\bibitem [{\citenamefont {Cancellieri}\ \emph {et~al.}(2013)\citenamefont
  {Cancellieri}, \citenamefont {Reinle-Schmitt}, \citenamefont {Kobayashi},
  \citenamefont {Strocov}, \citenamefont {Schmitt}, \citenamefont {Willmott},
  \citenamefont {Gariglio},\ and\ \citenamefont {Triscone}}]{r24}%
  \BibitemOpen
  \bibfield  {author} {\bibinfo {author} {\bibfnamefont {C.}~\bibnamefont
  {Cancellieri}}, \bibinfo {author} {\bibfnamefont {M.~L.}\ \bibnamefont
  {Reinle-Schmitt}}, \bibinfo {author} {\bibfnamefont {M.}~\bibnamefont
  {Kobayashi}}, \bibinfo {author} {\bibfnamefont {V.~N.}\ \bibnamefont
  {Strocov}}, \bibinfo {author} {\bibfnamefont {T.}~\bibnamefont {Schmitt}},
  \bibinfo {author} {\bibfnamefont {P.~R.}\ \bibnamefont {Willmott}}, \bibinfo
  {author} {\bibfnamefont {S.}~\bibnamefont {Gariglio}},\ and\ \bibinfo
  {author} {\bibfnamefont {J.~M.}\ \bibnamefont {Triscone}},\ }\href
  {https://doi.org/10.1103/PhysRevLett.110.137601} {\bibfield  {journal}
  {\bibinfo  {journal} {Phys. Rev. Lett.}\ }\textbf {\bibinfo {volume} {110}},\
  \bibinfo {pages} {137601} (\bibinfo {year} {2013})}\BibitemShut {NoStop}%
\bibitem [{\citenamefont {Peng}\ \emph {et~al.}(2020)\citenamefont {Peng},
  \citenamefont {Zou}, \citenamefont {Han}, \citenamefont {Albright},
  \citenamefont {Hong}, \citenamefont {Lau}, \citenamefont {Xu}, \citenamefont
  {Zhu}, \citenamefont {Walker},\ and\ \citenamefont {Ahn}}]{r25}%
  \BibitemOpen
  \bibfield  {author} {\bibinfo {author} {\bibfnamefont {R.}~\bibnamefont
  {Peng}}, \bibinfo {author} {\bibfnamefont {K.}~\bibnamefont {Zou}}, \bibinfo
  {author} {\bibfnamefont {M.~G.}\ \bibnamefont {Han}}, \bibinfo {author}
  {\bibfnamefont {S.~D.}\ \bibnamefont {Albright}}, \bibinfo {author}
  {\bibfnamefont {H.}~\bibnamefont {Hong}}, \bibinfo {author} {\bibfnamefont
  {C.}~\bibnamefont {Lau}}, \bibinfo {author} {\bibfnamefont {H.~C.}\
  \bibnamefont {Xu}}, \bibinfo {author} {\bibfnamefont {Y.}~\bibnamefont
  {Zhu}}, \bibinfo {author} {\bibfnamefont {F.~J.}\ \bibnamefont {Walker}},\
  and\ \bibinfo {author} {\bibfnamefont {C.~H.}\ \bibnamefont {Ahn}},\ }\href
  {https://doi.org/10.1126/sciadv.aay4517} {\bibfield  {journal} {\bibinfo
  {journal} {Sci. Adv.}\ }\textbf {\bibinfo {volume} {6}},\ \bibinfo {pages}
  {eaay4517} (\bibinfo {year} {2020})}\BibitemShut {NoStop}%
\bibitem [{\citenamefont {Song}\ \emph {et~al.}(2019)\citenamefont {Song},
  \citenamefont {Yu}, \citenamefont {Lou}, \citenamefont {Xie}, \citenamefont
  {Xu}, \citenamefont {Wen}, \citenamefont {Yao}, \citenamefont {Zhang},
  \citenamefont {Zhu}, \citenamefont {Guo}, \citenamefont {Peng},\ and\
  \citenamefont {Feng}}]{r26}%
  \BibitemOpen
  \bibfield  {author} {\bibinfo {author} {\bibfnamefont {Q.}~\bibnamefont
  {Song}}, \bibinfo {author} {\bibfnamefont {T.~L.}\ \bibnamefont {Yu}},
  \bibinfo {author} {\bibfnamefont {X.}~\bibnamefont {Lou}}, \bibinfo {author}
  {\bibfnamefont {B.~P.}\ \bibnamefont {Xie}}, \bibinfo {author} {\bibfnamefont
  {H.~C.}\ \bibnamefont {Xu}}, \bibinfo {author} {\bibfnamefont {C.~H.~P.}\
  \bibnamefont {Wen}}, \bibinfo {author} {\bibfnamefont {Q.}~\bibnamefont
  {Yao}}, \bibinfo {author} {\bibfnamefont {S.~Y.}\ \bibnamefont {Zhang}},
  \bibinfo {author} {\bibfnamefont {X.~T.}\ \bibnamefont {Zhu}}, \bibinfo
  {author} {\bibfnamefont {J.~D.}\ \bibnamefont {Guo}}, \bibinfo {author}
  {\bibfnamefont {R.}~\bibnamefont {Peng}},\ and\ \bibinfo {author}
  {\bibfnamefont {D.~L.}\ \bibnamefont {Feng}},\ }\href
  {https://doi.org/10.1038/s41467-019-08560-z} {\bibfield  {journal} {\bibinfo
  {journal} {Nat. Commun.}\ }\textbf {\bibinfo {volume} {10}},\ \bibinfo
  {pages} {758} (\bibinfo {year} {2019})}\BibitemShut {NoStop}%
\bibitem [{\citenamefont {Li}\ \emph {et~al.}(2014)\citenamefont {Li},
  \citenamefont {Peng}, \citenamefont {Zhang}, \citenamefont {Zhang},
  \citenamefont {Ding}, \citenamefont {Deng}, \citenamefont {Chang},
  \citenamefont {Song}, \citenamefont {Ji}, \citenamefont {Wang}, \citenamefont
  {He}, \citenamefont {Chen}, \citenamefont {Xue},\ and\ \citenamefont
  {Ma}}]{r27}%
  \BibitemOpen
  \bibfield  {author} {\bibinfo {author} {\bibfnamefont {Z.}~\bibnamefont
  {Li}}, \bibinfo {author} {\bibfnamefont {J.~P.}\ \bibnamefont {Peng}},
  \bibinfo {author} {\bibfnamefont {H.~M.}\ \bibnamefont {Zhang}}, \bibinfo
  {author} {\bibfnamefont {W.~H.}\ \bibnamefont {Zhang}}, \bibinfo {author}
  {\bibfnamefont {H.}~\bibnamefont {Ding}}, \bibinfo {author} {\bibfnamefont
  {P.}~\bibnamefont {Deng}}, \bibinfo {author} {\bibfnamefont {K.}~\bibnamefont
  {Chang}}, \bibinfo {author} {\bibfnamefont {C.~L.}\ \bibnamefont {Song}},
  \bibinfo {author} {\bibfnamefont {S.~H.}\ \bibnamefont {Ji}}, \bibinfo
  {author} {\bibfnamefont {L.}~\bibnamefont {Wang}}, \bibinfo {author}
  {\bibfnamefont {K.}~\bibnamefont {He}}, \bibinfo {author} {\bibfnamefont
  {X.}~\bibnamefont {Chen}}, \bibinfo {author} {\bibfnamefont {Q.~K.}\
  \bibnamefont {Xue}},\ and\ \bibinfo {author} {\bibfnamefont {X.~C.}\
  \bibnamefont {Ma}},\ }\href {https://doi.org/10.1088/0953-8984/26/26/265002}
  {\bibfield  {journal} {\bibinfo  {journal} {J. Phys.: Condens. Matter}\
  }\textbf {\bibinfo {volume} {26}},\ \bibinfo {pages} {265002} (\bibinfo
  {year} {2014})}\BibitemShut {NoStop}%
\bibitem [{\citenamefont {Gastiasoro}\ \emph {et~al.}(2022)\citenamefont
  {Gastiasoro}, \citenamefont {Temperini}, \citenamefont {Barone},\ and\
  \citenamefont {Lorenzana}}]{r28}%
  \BibitemOpen
  \bibfield  {author} {\bibinfo {author} {\bibfnamefont {M.~N.}\ \bibnamefont
  {Gastiasoro}}, \bibinfo {author} {\bibfnamefont {M.~E.}\ \bibnamefont
  {Temperini}}, \bibinfo {author} {\bibfnamefont {P.}~\bibnamefont {Barone}},\
  and\ \bibinfo {author} {\bibfnamefont {J.}~\bibnamefont {Lorenzana}},\ }\href
  {https://doi.org/10.1103/PhysRevB.105.224503} {\bibfield  {journal} {\bibinfo
   {journal} {Phys. Rev. B}\ }\textbf {\bibinfo {volume} {105}},\ \bibinfo
  {pages} {224503} (\bibinfo {year} {2022})}\BibitemShut {NoStop}%
\bibitem [{\citenamefont {Sun}\ \emph {et~al.}(2021)\citenamefont {Sun},
  \citenamefont {Liu}, \citenamefont {Hong}, \citenamefont {Chen},
  \citenamefont {Zhang},\ and\ \citenamefont {Xie}}]{r29}%
  \BibitemOpen
  \bibfield  {author} {\bibinfo {author} {\bibfnamefont {Y.}~\bibnamefont
  {Sun}}, \bibinfo {author} {\bibfnamefont {Y.}~\bibnamefont {Liu}}, \bibinfo
  {author} {\bibfnamefont {S.}~\bibnamefont {Hong}}, \bibinfo {author}
  {\bibfnamefont {Z.}~\bibnamefont {Chen}}, \bibinfo {author} {\bibfnamefont
  {M.}~\bibnamefont {Zhang}},\ and\ \bibinfo {author} {\bibfnamefont
  {Y.}~\bibnamefont {Xie}},\ }\href
  {https://doi.org/10.1103/PhysRevLett.127.086804} {\bibfield  {journal}
  {\bibinfo  {journal} {Phys. Rev. Lett.}\ }\textbf {\bibinfo {volume} {127}},\
  \bibinfo {pages} {086804} (\bibinfo {year} {2021})}\BibitemShut {NoStop}%
\bibitem [{\citenamefont {Strocov}\ \emph {et~al.}(2014)\citenamefont
  {Strocov}, \citenamefont {Wang}, \citenamefont {Shi}, \citenamefont
  {Kobayashi}, \citenamefont {Krempasky}, \citenamefont {Hess}, \citenamefont
  {Schmitt},\ and\ \citenamefont {Patthey}}]{r30}%
  \BibitemOpen
  \bibfield  {author} {\bibinfo {author} {\bibfnamefont {V.~N.}\ \bibnamefont
  {Strocov}}, \bibinfo {author} {\bibfnamefont {X.}~\bibnamefont {Wang}},
  \bibinfo {author} {\bibfnamefont {M.}~\bibnamefont {Shi}}, \bibinfo {author}
  {\bibfnamefont {M.}~\bibnamefont {Kobayashi}}, \bibinfo {author}
  {\bibfnamefont {J.}~\bibnamefont {Krempasky}}, \bibinfo {author}
  {\bibfnamefont {C.}~\bibnamefont {Hess}}, \bibinfo {author} {\bibfnamefont
  {T.}~\bibnamefont {Schmitt}},\ and\ \bibinfo {author} {\bibfnamefont
  {L.}~\bibnamefont {Patthey}},\ }\href
  {https://doi.org/10.1107/S1600577513019085} {\bibfield  {journal} {\bibinfo
  {journal} {J. Synchrotron. Rad.}\ }\textbf {\bibinfo {volume} {21}},\
  \bibinfo {pages} {32} (\bibinfo {year} {2014})}\BibitemShut {NoStop}%
\bibitem [{\citenamefont {Strocov}\ \emph {et~al.}(2010)\citenamefont
  {Strocov}, \citenamefont {Schmitt}, \citenamefont {Flechsig}, \citenamefont
  {Schmidt}, \citenamefont {Imhof}, \citenamefont {Chen}, \citenamefont
  {Raabe}, \citenamefont {Betemps}, \citenamefont {Zimoch}, \citenamefont
  {Krempasky}, \citenamefont {Wang}, \citenamefont {Grioni}, \citenamefont
  {Piazzalunga},\ and\ \citenamefont {Patthey}}]{r31}%
  \BibitemOpen
  \bibfield  {author} {\bibinfo {author} {\bibfnamefont {V.~N.}\ \bibnamefont
  {Strocov}}, \bibinfo {author} {\bibfnamefont {T.}~\bibnamefont {Schmitt}},
  \bibinfo {author} {\bibfnamefont {U.}~\bibnamefont {Flechsig}}, \bibinfo
  {author} {\bibfnamefont {T.}~\bibnamefont {Schmidt}}, \bibinfo {author}
  {\bibfnamefont {A.}~\bibnamefont {Imhof}}, \bibinfo {author} {\bibfnamefont
  {Q.}~\bibnamefont {Chen}}, \bibinfo {author} {\bibfnamefont {J.}~\bibnamefont
  {Raabe}}, \bibinfo {author} {\bibfnamefont {R.}~\bibnamefont {Betemps}},
  \bibinfo {author} {\bibfnamefont {D.}~\bibnamefont {Zimoch}}, \bibinfo
  {author} {\bibfnamefont {J.}~\bibnamefont {Krempasky}}, \bibinfo {author}
  {\bibfnamefont {X.}~\bibnamefont {Wang}}, \bibinfo {author} {\bibfnamefont
  {M.}~\bibnamefont {Grioni}}, \bibinfo {author} {\bibfnamefont
  {A.}~\bibnamefont {Piazzalunga}},\ and\ \bibinfo {author} {\bibfnamefont
  {L.}~\bibnamefont {Patthey}},\ }\href
  {https://doi.org/10.1107/S0909049510019862} {\bibfield  {journal} {\bibinfo
  {journal} {J. Synchrotron. Rad.}\ }\textbf {\bibinfo {volume} {17}},\
  \bibinfo {pages} {631} (\bibinfo {year} {2010})}\BibitemShut {NoStop}%
\bibitem [{\citenamefont {Zhang}\ \emph {et~al.}(2019)\citenamefont {Zhang},
  \citenamefont {Yan}, \citenamefont {Zhang}, \citenamefont {Wang},
  \citenamefont {Xiong}, \citenamefont {Zhang}, \citenamefont {Qi},
  \citenamefont {Zhang}, \citenamefont {Han}, \citenamefont {Wu}, \citenamefont
  {Liu}, \citenamefont {Chen}, \citenamefont {Shen},\ and\ \citenamefont
  {Sun}}]{r32}%
  \BibitemOpen
  \bibfield  {author} {\bibinfo {author} {\bibfnamefont {H.}~\bibnamefont
  {Zhang}}, \bibinfo {author} {\bibfnamefont {X.}~\bibnamefont {Yan}}, \bibinfo
  {author} {\bibfnamefont {X.}~\bibnamefont {Zhang}}, \bibinfo {author}
  {\bibfnamefont {S.}~\bibnamefont {Wang}}, \bibinfo {author} {\bibfnamefont
  {C.}~\bibnamefont {Xiong}}, \bibinfo {author} {\bibfnamefont
  {H.}~\bibnamefont {Zhang}}, \bibinfo {author} {\bibfnamefont
  {S.}~\bibnamefont {Qi}}, \bibinfo {author} {\bibfnamefont {J.}~\bibnamefont
  {Zhang}}, \bibinfo {author} {\bibfnamefont {F.}~\bibnamefont {Han}}, \bibinfo
  {author} {\bibfnamefont {N.}~\bibnamefont {Wu}}, \bibinfo {author}
  {\bibfnamefont {B.}~\bibnamefont {Liu}}, \bibinfo {author} {\bibfnamefont
  {Y.}~\bibnamefont {Chen}}, \bibinfo {author} {\bibfnamefont {B.}~\bibnamefont
  {Shen}},\ and\ \bibinfo {author} {\bibfnamefont {J.}~\bibnamefont {Sun}},\
  }\href {https://doi.org/10.1021/acsnano.8b07622} {\bibfield  {journal}
  {\bibinfo  {journal} {ACS Nano}\ }\textbf {\bibinfo {volume} {13}},\ \bibinfo
  {pages} {609} (\bibinfo {year} {2019})}\BibitemShut {NoStop}%
\bibitem [{\citenamefont {Santander-Syro}\ \emph {et~al.}(2012)\citenamefont
  {Santander-Syro}, \citenamefont {Bareille}, \citenamefont {Fortuna},
  \citenamefont {Copie}, \citenamefont {Gabay}, \citenamefont {Bertran},
  \citenamefont {Taleb-Ibrahimi}, \citenamefont {Le~Fèvre}, \citenamefont
  {Herranz}, \citenamefont {Reyren}, \citenamefont {Bibes}, \citenamefont
  {Barthélémy}, \citenamefont {Lecoeur}, \citenamefont {Guevara},\ and\
  \citenamefont {Rozenberg}}]{r33}%
  \BibitemOpen
  \bibfield  {author} {\bibinfo {author} {\bibfnamefont {A.~F.}\ \bibnamefont
  {Santander-Syro}}, \bibinfo {author} {\bibfnamefont {C.}~\bibnamefont
  {Bareille}}, \bibinfo {author} {\bibfnamefont {F.}~\bibnamefont {Fortuna}},
  \bibinfo {author} {\bibfnamefont {O.}~\bibnamefont {Copie}}, \bibinfo
  {author} {\bibfnamefont {M.}~\bibnamefont {Gabay}}, \bibinfo {author}
  {\bibfnamefont {F.}~\bibnamefont {Bertran}}, \bibinfo {author} {\bibfnamefont
  {A.}~\bibnamefont {Taleb-Ibrahimi}}, \bibinfo {author} {\bibfnamefont
  {P.}~\bibnamefont {Le~Fèvre}}, \bibinfo {author} {\bibfnamefont
  {G.}~\bibnamefont {Herranz}}, \bibinfo {author} {\bibfnamefont
  {N.}~\bibnamefont {Reyren}}, \bibinfo {author} {\bibfnamefont
  {M.}~\bibnamefont {Bibes}}, \bibinfo {author} {\bibfnamefont
  {A.}~\bibnamefont {Barthélémy}}, \bibinfo {author} {\bibfnamefont
  {P.}~\bibnamefont {Lecoeur}}, \bibinfo {author} {\bibfnamefont
  {J.}~\bibnamefont {Guevara}},\ and\ \bibinfo {author} {\bibfnamefont {M.~J.}\
  \bibnamefont {Rozenberg}},\ }\href
  {https://doi.org/10.1103/PhysRevB.86.121107} {\bibfield  {journal} {\bibinfo
  {journal} {Phys. Rev. B}\ }\textbf {\bibinfo {volume} {86}},\ \bibinfo
  {pages} {121107} (\bibinfo {year} {2012})}\BibitemShut {NoStop}%
\bibitem [{\citenamefont {King}\ \emph {et~al.}(2012)\citenamefont {King},
  \citenamefont {He}, \citenamefont {Eknapakul}, \citenamefont {Buaphet},
  \citenamefont {Mo}, \citenamefont {Kaneko}, \citenamefont {Harashima},
  \citenamefont {Hikita}, \citenamefont {Bahramy}, \citenamefont {Bell},
  \citenamefont {Hussain}, \citenamefont {Tokura}, \citenamefont {Shen},
  \citenamefont {Hwang}, \citenamefont {Baumberger},\ and\ \citenamefont
  {Meevasana}}]{r34}%
  \BibitemOpen
  \bibfield  {author} {\bibinfo {author} {\bibfnamefont {P.~D.}\ \bibnamefont
  {King}}, \bibinfo {author} {\bibfnamefont {R.~H.}\ \bibnamefont {He}},
  \bibinfo {author} {\bibfnamefont {T.}~\bibnamefont {Eknapakul}}, \bibinfo
  {author} {\bibfnamefont {P.}~\bibnamefont {Buaphet}}, \bibinfo {author}
  {\bibfnamefont {S.~K.}\ \bibnamefont {Mo}}, \bibinfo {author} {\bibfnamefont
  {Y.}~\bibnamefont {Kaneko}}, \bibinfo {author} {\bibfnamefont
  {S.}~\bibnamefont {Harashima}}, \bibinfo {author} {\bibfnamefont
  {Y.}~\bibnamefont {Hikita}}, \bibinfo {author} {\bibfnamefont {M.~S.}\
  \bibnamefont {Bahramy}}, \bibinfo {author} {\bibfnamefont {C.}~\bibnamefont
  {Bell}}, \bibinfo {author} {\bibfnamefont {Z.}~\bibnamefont {Hussain}},
  \bibinfo {author} {\bibfnamefont {Y.}~\bibnamefont {Tokura}}, \bibinfo
  {author} {\bibfnamefont {Z.~X.}\ \bibnamefont {Shen}}, \bibinfo {author}
  {\bibfnamefont {H.~Y.}\ \bibnamefont {Hwang}}, \bibinfo {author}
  {\bibfnamefont {F.}~\bibnamefont {Baumberger}},\ and\ \bibinfo {author}
  {\bibfnamefont {W.}~\bibnamefont {Meevasana}},\ }\href
  {https://doi.org/10.1103/PhysRevLett.108.117602} {\bibfield  {journal}
  {\bibinfo  {journal} {Phys. Rev. Lett.}\ }\textbf {\bibinfo {volume} {108}},\
  \bibinfo {pages} {117602} (\bibinfo {year} {2012})}\BibitemShut {NoStop}%
\bibitem [{\citenamefont {Bareille}\ \emph {et~al.}(2014)\citenamefont
  {Bareille}, \citenamefont {Fortuna}, \citenamefont {Rodel}, \citenamefont
  {Bertran}, \citenamefont {Gabay}, \citenamefont {Cubelos}, \citenamefont
  {Taleb-Ibrahimi}, \citenamefont {Le~Fevre}, \citenamefont {Bibes},
  \citenamefont {Barthelemy}, \citenamefont {Maroutian}, \citenamefont
  {Lecoeur}, \citenamefont {Rozenberg},\ and\ \citenamefont
  {Santander-Syro}}]{r35}%
  \BibitemOpen
  \bibfield  {author} {\bibinfo {author} {\bibfnamefont {C.}~\bibnamefont
  {Bareille}}, \bibinfo {author} {\bibfnamefont {F.}~\bibnamefont {Fortuna}},
  \bibinfo {author} {\bibfnamefont {T.~C.}\ \bibnamefont {Rodel}}, \bibinfo
  {author} {\bibfnamefont {F.}~\bibnamefont {Bertran}}, \bibinfo {author}
  {\bibfnamefont {M.}~\bibnamefont {Gabay}}, \bibinfo {author} {\bibfnamefont
  {O.~H.}\ \bibnamefont {Cubelos}}, \bibinfo {author} {\bibfnamefont
  {A.}~\bibnamefont {Taleb-Ibrahimi}}, \bibinfo {author} {\bibfnamefont
  {P.}~\bibnamefont {Le~Fevre}}, \bibinfo {author} {\bibfnamefont
  {M.}~\bibnamefont {Bibes}}, \bibinfo {author} {\bibfnamefont
  {A.}~\bibnamefont {Barthelemy}}, \bibinfo {author} {\bibfnamefont
  {T.}~\bibnamefont {Maroutian}}, \bibinfo {author} {\bibfnamefont
  {P.}~\bibnamefont {Lecoeur}}, \bibinfo {author} {\bibfnamefont {M.~J.}\
  \bibnamefont {Rozenberg}},\ and\ \bibinfo {author} {\bibfnamefont {A.~F.}\
  \bibnamefont {Santander-Syro}},\ }\href {https://doi.org/10.1038/srep03586}
  {\bibfield  {journal} {\bibinfo  {journal} {Sci. Rep.}\ }\textbf {\bibinfo
  {volume} {4}},\ \bibinfo {pages} {3586} (\bibinfo {year} {2014})}\BibitemShut
  {NoStop}%
\bibitem [{\citenamefont {Bruno}\ \emph {et~al.}(2019)\citenamefont {Bruno},
  \citenamefont {McKeown~Walker}, \citenamefont {Riccò}, \citenamefont
  {la~Torre}, \citenamefont {Wang}, \citenamefont {Tamai}, \citenamefont {Kim},
  \citenamefont {Hoesch}, \citenamefont {Bahramy},\ and\ \citenamefont
  {Baumberger}}]{r36}%
  \BibitemOpen
  \bibfield  {author} {\bibinfo {author} {\bibfnamefont {F.~Y.}\ \bibnamefont
  {Bruno}}, \bibinfo {author} {\bibfnamefont {S.}~\bibnamefont
  {McKeown~Walker}}, \bibinfo {author} {\bibfnamefont {S.}~\bibnamefont
  {Riccò}}, \bibinfo {author} {\bibfnamefont {A.}~\bibnamefont {la~Torre}},
  \bibinfo {author} {\bibfnamefont {Z.}~\bibnamefont {Wang}}, \bibinfo {author}
  {\bibfnamefont {A.}~\bibnamefont {Tamai}}, \bibinfo {author} {\bibfnamefont
  {T.~K.}\ \bibnamefont {Kim}}, \bibinfo {author} {\bibfnamefont
  {M.}~\bibnamefont {Hoesch}}, \bibinfo {author} {\bibfnamefont {M.~S.}\
  \bibnamefont {Bahramy}},\ and\ \bibinfo {author} {\bibfnamefont
  {F.}~\bibnamefont {Baumberger}},\ }\href
  {https://doi.org/10.1002/aelm.201800860} {\bibfield  {journal} {\bibinfo
  {journal} {Adv. Electron. Mater.}\ }\textbf {\bibinfo {volume} {5}},\
  \bibinfo {pages} {1800860} (\bibinfo {year} {2019})}\BibitemShut {NoStop}%
\bibitem [{\citenamefont {Moser}(2017)}]{r45}%
  \BibitemOpen
  \bibfield  {author} {\bibinfo {author} {\bibfnamefont {S.}~\bibnamefont
  {Moser}},\ }\href {https://doi.org/10.1016/j.elspec.2016.11.007} {\bibfield
  {journal} {\bibinfo  {journal} {Journal of Electron Spectroscopy and Related
  Phenomena}\ }\textbf {\bibinfo {volume} {214}},\ \bibinfo {pages} {29}
  (\bibinfo {year} {2017})}\BibitemShut {NoStop}%
\bibitem [{\citenamefont {Strocov}\ \emph {et~al.}(2012)\citenamefont
  {Strocov}, \citenamefont {Shi}, \citenamefont {Kobayashi}, \citenamefont
  {Monney}, \citenamefont {Wang}, \citenamefont {Krempasky}, \citenamefont
  {Schmitt}, \citenamefont {Patthey}, \citenamefont {Berger},\ and\
  \citenamefont {Blaha}}]{r37}%
  \BibitemOpen
  \bibfield  {author} {\bibinfo {author} {\bibfnamefont {V.~N.}\ \bibnamefont
  {Strocov}}, \bibinfo {author} {\bibfnamefont {M.}~\bibnamefont {Shi}},
  \bibinfo {author} {\bibfnamefont {M.}~\bibnamefont {Kobayashi}}, \bibinfo
  {author} {\bibfnamefont {C.}~\bibnamefont {Monney}}, \bibinfo {author}
  {\bibfnamefont {X.}~\bibnamefont {Wang}}, \bibinfo {author} {\bibfnamefont
  {J.}~\bibnamefont {Krempasky}}, \bibinfo {author} {\bibfnamefont
  {T.}~\bibnamefont {Schmitt}}, \bibinfo {author} {\bibfnamefont
  {L.}~\bibnamefont {Patthey}}, \bibinfo {author} {\bibfnamefont
  {H.}~\bibnamefont {Berger}},\ and\ \bibinfo {author} {\bibfnamefont
  {P.}~\bibnamefont {Blaha}},\ }\href
  {https://doi.org/10.1103/PhysRevLett.109.086401} {\bibfield  {journal}
  {\bibinfo  {journal} {Phys. Rev. Lett.}\ }\textbf {\bibinfo {volume} {109}},\
  \bibinfo {pages} {086401} (\bibinfo {year} {2012})}\BibitemShut {NoStop}%
\bibitem [{\citenamefont {Luttinger}(1960)}]{r38}%
  \BibitemOpen
  \bibfield  {author} {\bibinfo {author} {\bibfnamefont {J.~M.}\ \bibnamefont
  {Luttinger}},\ }\href {https://doi.org/10.1103/PhysRev.119.1153} {\bibfield
  {journal} {\bibinfo  {journal} {Phys. Rev.}\ }\textbf {\bibinfo {volume}
  {119}},\ \bibinfo {pages} {1153} (\bibinfo {year} {1960})}\BibitemShut
  {NoStop}%
\bibitem [{\citenamefont {Scopigno}\ \emph {et~al.}(2016)\citenamefont
  {Scopigno}, \citenamefont {Bucheli}, \citenamefont {Caprara}, \citenamefont
  {Biscaras}, \citenamefont {Bergeal}, \citenamefont {Lesueur},\ and\
  \citenamefont {Grilli}}]{r39}%
  \BibitemOpen
  \bibfield  {author} {\bibinfo {author} {\bibfnamefont {N.}~\bibnamefont
  {Scopigno}}, \bibinfo {author} {\bibfnamefont {D.}~\bibnamefont {Bucheli}},
  \bibinfo {author} {\bibfnamefont {S.}~\bibnamefont {Caprara}}, \bibinfo
  {author} {\bibfnamefont {J.}~\bibnamefont {Biscaras}}, \bibinfo {author}
  {\bibfnamefont {N.}~\bibnamefont {Bergeal}}, \bibinfo {author} {\bibfnamefont
  {J.}~\bibnamefont {Lesueur}},\ and\ \bibinfo {author} {\bibfnamefont
  {M.}~\bibnamefont {Grilli}},\ }\href
  {https://doi.org/10.1103/PhysRevLett.116.026804} {\bibfield  {journal}
  {\bibinfo  {journal} {Phys. Rev. Lett.}\ }\textbf {\bibinfo {volume} {116}},\
  \bibinfo {pages} {026804} (\bibinfo {year} {2016})}\BibitemShut {NoStop}%
\bibitem [{\citenamefont {Strocov}\ \emph {et~al.}(2019)\citenamefont
  {Strocov}, \citenamefont {Chikina}, \citenamefont {Caputo}, \citenamefont
  {Husanu}, \citenamefont {Bisti}, \citenamefont {Bracher}, \citenamefont
  {Schmitt}, \citenamefont {Miletto~Granozio}, \citenamefont {Vaz},\ and\
  \citenamefont {Lechermann}}]{r40}%
  \BibitemOpen
  \bibfield  {author} {\bibinfo {author} {\bibfnamefont {V.~N.}\ \bibnamefont
  {Strocov}}, \bibinfo {author} {\bibfnamefont {A.}~\bibnamefont {Chikina}},
  \bibinfo {author} {\bibfnamefont {M.}~\bibnamefont {Caputo}}, \bibinfo
  {author} {\bibfnamefont {M.~A.}\ \bibnamefont {Husanu}}, \bibinfo {author}
  {\bibfnamefont {F.}~\bibnamefont {Bisti}}, \bibinfo {author} {\bibfnamefont
  {D.}~\bibnamefont {Bracher}}, \bibinfo {author} {\bibfnamefont
  {T.}~\bibnamefont {Schmitt}}, \bibinfo {author} {\bibfnamefont
  {F.}~\bibnamefont {Miletto~Granozio}}, \bibinfo {author} {\bibfnamefont
  {C.~A.~F.}\ \bibnamefont {Vaz}},\ and\ \bibinfo {author} {\bibfnamefont
  {F.}~\bibnamefont {Lechermann}},\ }\href
  {https://doi.org/10.1103/PhysRevMaterials.3.106001} {\bibfield  {journal}
  {\bibinfo  {journal} {Phys. Rev. Materials}\ }\textbf {\bibinfo {volume}
  {3}},\ \bibinfo {pages} {106001} (\bibinfo {year} {2019})}\BibitemShut
  {NoStop}%
\bibitem [{\citenamefont {Kozuka}\ \emph {et~al.}(2009)\citenamefont {Kozuka},
  \citenamefont {Kim}, \citenamefont {Bell}, \citenamefont {Kim}, \citenamefont
  {Hikita},\ and\ \citenamefont {Hwang}}]{r41}%
  \BibitemOpen
  \bibfield  {author} {\bibinfo {author} {\bibfnamefont {Y.}~\bibnamefont
  {Kozuka}}, \bibinfo {author} {\bibfnamefont {M.}~\bibnamefont {Kim}},
  \bibinfo {author} {\bibfnamefont {C.}~\bibnamefont {Bell}}, \bibinfo {author}
  {\bibfnamefont {B.~G.}\ \bibnamefont {Kim}}, \bibinfo {author} {\bibfnamefont
  {Y.}~\bibnamefont {Hikita}},\ and\ \bibinfo {author} {\bibfnamefont {H.~Y.}\
  \bibnamefont {Hwang}},\ }\href {https://doi.org/10.1038/nature08566}
  {\bibfield  {journal} {\bibinfo  {journal} {Nature}\ }\textbf {\bibinfo
  {volume} {462}},\ \bibinfo {pages} {487} (\bibinfo {year}
  {2009})}\BibitemShut {NoStop}%
\bibitem [{\citenamefont {Santander-Syro}\ \emph {et~al.}(2011)\citenamefont
  {Santander-Syro}, \citenamefont {Copie}, \citenamefont {Kondo}, \citenamefont
  {Fortuna}, \citenamefont {Pailhes}, \citenamefont {Weht}, \citenamefont
  {Qiu}, \citenamefont {Bertran}, \citenamefont {Nicolaou}, \citenamefont
  {Taleb-Ibrahimi}, \citenamefont {Le~Fevre}, \citenamefont {Herranz},
  \citenamefont {Bibes}, \citenamefont {Reyren}, \citenamefont {Apertet},
  \citenamefont {Lecoeur}, \citenamefont {Barthelemy},\ and\ \citenamefont
  {Rozenberg}}]{r42}%
  \BibitemOpen
  \bibfield  {author} {\bibinfo {author} {\bibfnamefont {A.~F.}\ \bibnamefont
  {Santander-Syro}}, \bibinfo {author} {\bibfnamefont {O.}~\bibnamefont
  {Copie}}, \bibinfo {author} {\bibfnamefont {T.}~\bibnamefont {Kondo}},
  \bibinfo {author} {\bibfnamefont {F.}~\bibnamefont {Fortuna}}, \bibinfo
  {author} {\bibfnamefont {S.}~\bibnamefont {Pailhes}}, \bibinfo {author}
  {\bibfnamefont {R.}~\bibnamefont {Weht}}, \bibinfo {author} {\bibfnamefont
  {X.~G.}\ \bibnamefont {Qiu}}, \bibinfo {author} {\bibfnamefont
  {F.}~\bibnamefont {Bertran}}, \bibinfo {author} {\bibfnamefont
  {A.}~\bibnamefont {Nicolaou}}, \bibinfo {author} {\bibfnamefont
  {A.}~\bibnamefont {Taleb-Ibrahimi}}, \bibinfo {author} {\bibfnamefont
  {P.}~\bibnamefont {Le~Fevre}}, \bibinfo {author} {\bibfnamefont
  {G.}~\bibnamefont {Herranz}}, \bibinfo {author} {\bibfnamefont
  {M.}~\bibnamefont {Bibes}}, \bibinfo {author} {\bibfnamefont
  {N.}~\bibnamefont {Reyren}}, \bibinfo {author} {\bibfnamefont
  {Y.}~\bibnamefont {Apertet}}, \bibinfo {author} {\bibfnamefont
  {P.}~\bibnamefont {Lecoeur}}, \bibinfo {author} {\bibfnamefont
  {A.}~\bibnamefont {Barthelemy}},\ and\ \bibinfo {author} {\bibfnamefont
  {M.~J.}\ \bibnamefont {Rozenberg}},\ }\href
  {https://doi.org/10.1038/nature09720} {\bibfield  {journal} {\bibinfo
  {journal} {Nature}\ }\textbf {\bibinfo {volume} {469}},\ \bibinfo {pages}
  {189} (\bibinfo {year} {2011})}\BibitemShut {NoStop}%
\bibitem [{\citenamefont {Zhang}\ \emph {et~al.}(2010)\citenamefont {Zhang},
  \citenamefont {Yang}, \citenamefont {Chen}, \citenamefont {Zhou},
  \citenamefont {Wang}, \citenamefont {Chen}, \citenamefont {Arita},
  \citenamefont {Shimada}, \citenamefont {Namatame}, \citenamefont {Taniguchi},
  \citenamefont {Hu}, \citenamefont {Xie},\ and\ \citenamefont {Feng}}]{r43}%
  \BibitemOpen
  \bibfield  {author} {\bibinfo {author} {\bibfnamefont {Y.}~\bibnamefont
  {Zhang}}, \bibinfo {author} {\bibfnamefont {L.~X.}\ \bibnamefont {Yang}},
  \bibinfo {author} {\bibfnamefont {F.}~\bibnamefont {Chen}}, \bibinfo {author}
  {\bibfnamefont {B.}~\bibnamefont {Zhou}}, \bibinfo {author} {\bibfnamefont
  {X.~F.}\ \bibnamefont {Wang}}, \bibinfo {author} {\bibfnamefont {X.~H.}\
  \bibnamefont {Chen}}, \bibinfo {author} {\bibfnamefont {M.}~\bibnamefont
  {Arita}}, \bibinfo {author} {\bibfnamefont {K.}~\bibnamefont {Shimada}},
  \bibinfo {author} {\bibfnamefont {H.}~\bibnamefont {Namatame}}, \bibinfo
  {author} {\bibfnamefont {M.}~\bibnamefont {Taniguchi}}, \bibinfo {author}
  {\bibfnamefont {J.~P.}\ \bibnamefont {Hu}}, \bibinfo {author} {\bibfnamefont
  {B.~P.}\ \bibnamefont {Xie}},\ and\ \bibinfo {author} {\bibfnamefont {D.~L.}\
  \bibnamefont {Feng}},\ }\href
  {https://doi.org/10.1103/PhysRevLett.105.117003} {\bibfield  {journal}
  {\bibinfo  {journal} {Phys. Rev. Lett.}\ }\textbf {\bibinfo {volume} {105}},\
  \bibinfo {pages} {117003} (\bibinfo {year} {2010})}\BibitemShut {NoStop}%
\bibitem [{\citenamefont {Xu}\ \emph {et~al.}(2014)\citenamefont {Xu},
  \citenamefont {Xu}, \citenamefont {Peng}, \citenamefont {Zhang},
  \citenamefont {Ge}, \citenamefont {Qin}, \citenamefont {Xia}, \citenamefont
  {Ying}, \citenamefont {Chen}, \citenamefont {Yu}, \citenamefont {Zou},
  \citenamefont {Arita}, \citenamefont {Shimada}, \citenamefont {Taniguchi},
  \citenamefont {Lu}, \citenamefont {Xie},\ and\ \citenamefont {Feng}}]{r44}%
  \BibitemOpen
  \bibfield  {author} {\bibinfo {author} {\bibfnamefont {H.~C.}\ \bibnamefont
  {Xu}}, \bibinfo {author} {\bibfnamefont {M.}~\bibnamefont {Xu}}, \bibinfo
  {author} {\bibfnamefont {R.}~\bibnamefont {Peng}}, \bibinfo {author}
  {\bibfnamefont {Y.}~\bibnamefont {Zhang}}, \bibinfo {author} {\bibfnamefont
  {Q.~Q.}\ \bibnamefont {Ge}}, \bibinfo {author} {\bibfnamefont
  {F.}~\bibnamefont {Qin}}, \bibinfo {author} {\bibfnamefont {M.}~\bibnamefont
  {Xia}}, \bibinfo {author} {\bibfnamefont {J.~J.}\ \bibnamefont {Ying}},
  \bibinfo {author} {\bibfnamefont {X.~H.}\ \bibnamefont {Chen}}, \bibinfo
  {author} {\bibfnamefont {X.~L.}\ \bibnamefont {Yu}}, \bibinfo {author}
  {\bibfnamefont {L.~J.}\ \bibnamefont {Zou}}, \bibinfo {author} {\bibfnamefont
  {M.}~\bibnamefont {Arita}}, \bibinfo {author} {\bibfnamefont
  {K.}~\bibnamefont {Shimada}}, \bibinfo {author} {\bibfnamefont
  {M.}~\bibnamefont {Taniguchi}}, \bibinfo {author} {\bibfnamefont {D.~H.}\
  \bibnamefont {Lu}}, \bibinfo {author} {\bibfnamefont {B.~P.}\ \bibnamefont
  {Xie}},\ and\ \bibinfo {author} {\bibfnamefont {D.~L.}\ \bibnamefont
  {Feng}},\ }\href {https://doi.org/10.1103/PhysRevB.89.155108} {\bibfield
  {journal} {\bibinfo  {journal} {Phys. Rev. B}\ }\textbf {\bibinfo {volume}
  {89}},\ \bibinfo {pages} {155108} (\bibinfo {year} {2014})}\BibitemShut
  {NoStop}%
\bibitem [{\citenamefont {Tougaard}(1988)}]{r48}%
  \BibitemOpen
  \bibfield  {author} {\bibinfo {author} {\bibfnamefont {S.}~\bibnamefont
  {Tougaard}},\ }\href {https://doi.org/10.1002/sia.740110902} {\bibfield
  {journal} {\bibinfo  {journal} {Surface and Interface Analysis}\ }\textbf
  {\bibinfo {volume} {11}},\ \bibinfo {pages} {453} (\bibinfo {year}
  {1988})}\BibitemShut {NoStop}%
\bibitem [{\citenamefont {Tougaard}(2021)}]{r49}%
  \BibitemOpen
  \bibfield  {author} {\bibinfo {author} {\bibfnamefont {S.}~\bibnamefont
  {Tougaard}},\ }\href {https://doi.org/10.1116/6.0000661} {\bibfield
  {journal} {\bibinfo  {journal} {J. Vac. Sci. Technol. A}\ }\textbf {\bibinfo
  {volume} {39}},\ \bibinfo {pages} {011201} (\bibinfo {year}
  {2021})}\BibitemShut {NoStop}%
\bibitem [{\citenamefont {Liu}\ \emph {et~al.}(2021{\natexlab{b}})\citenamefont
  {Liu}, \citenamefont {Day}, \citenamefont {Li}, \citenamefont {Roemer},
  \citenamefont {Zhdanovich}, \citenamefont {Gorovikov}, \citenamefont
  {Pedersen}, \citenamefont {Jiang}, \citenamefont {Lee}, \citenamefont
  {Schneider}, \citenamefont {Wong}, \citenamefont {Dosanjh}, \citenamefont
  {Walker}, \citenamefont {Ahn}, \citenamefont {Levy}, \citenamefont
  {Damascelli}, \citenamefont {Sawatzky},\ and\ \citenamefont {Zou}}]{r50}%
  \BibitemOpen
  \bibfield  {author} {\bibinfo {author} {\bibfnamefont {C.}~\bibnamefont
  {Liu}}, \bibinfo {author} {\bibfnamefont {R.~P.}\ \bibnamefont {Day}},
  \bibinfo {author} {\bibfnamefont {F.}~\bibnamefont {Li}}, \bibinfo {author}
  {\bibfnamefont {R.~L.}\ \bibnamefont {Roemer}}, \bibinfo {author}
  {\bibfnamefont {S.}~\bibnamefont {Zhdanovich}}, \bibinfo {author}
  {\bibfnamefont {S.}~\bibnamefont {Gorovikov}}, \bibinfo {author}
  {\bibfnamefont {T.~M.}\ \bibnamefont {Pedersen}}, \bibinfo {author}
  {\bibfnamefont {J.}~\bibnamefont {Jiang}}, \bibinfo {author} {\bibfnamefont
  {S.}~\bibnamefont {Lee}}, \bibinfo {author} {\bibfnamefont {M.}~\bibnamefont
  {Schneider}}, \bibinfo {author} {\bibfnamefont {D.}~\bibnamefont {Wong}},
  \bibinfo {author} {\bibfnamefont {P.}~\bibnamefont {Dosanjh}}, \bibinfo
  {author} {\bibfnamefont {F.~J.}\ \bibnamefont {Walker}}, \bibinfo {author}
  {\bibfnamefont {C.~H.}\ \bibnamefont {Ahn}}, \bibinfo {author} {\bibfnamefont
  {G.}~\bibnamefont {Levy}}, \bibinfo {author} {\bibfnamefont {A.}~\bibnamefont
  {Damascelli}}, \bibinfo {author} {\bibfnamefont {G.~A.}\ \bibnamefont
  {Sawatzky}},\ and\ \bibinfo {author} {\bibfnamefont {K.}~\bibnamefont
  {Zou}},\ }\href {https://doi.org/10.1038/s41467-021-24783-5} {\bibfield
  {journal} {\bibinfo  {journal} {Nat. Commun.}\ }\textbf {\bibinfo {volume}
  {12}},\ \bibinfo {pages} {4573} (\bibinfo {year}
  {2021}{\natexlab{b}})}\BibitemShut {NoStop}%
\bibitem [{\citenamefont {Li}\ \emph {et~al.}(2019)\citenamefont {Li},
  \citenamefont {Devereaux},\ and\ \citenamefont {Lee}}]{r51}%
  \BibitemOpen
  \bibfield  {author} {\bibinfo {author} {\bibfnamefont {Z.-X.}\ \bibnamefont
  {Li}}, \bibinfo {author} {\bibfnamefont {T.~P.}\ \bibnamefont {Devereaux}},\
  and\ \bibinfo {author} {\bibfnamefont {D.-H.}\ \bibnamefont {Lee}},\ }\href
  {https://doi.org/10.1103/PhysRevB.100.241101} {\bibfield  {journal} {\bibinfo
   {journal} {Phys. Rev. B}\ }\textbf {\bibinfo {volume} {100}},\ \bibinfo
  {pages} {241101} (\bibinfo {year} {2019})}\BibitemShut {NoStop}%
\bibitem [{\citenamefont {Sobota}\ \emph {et~al.}(2021)\citenamefont {Sobota},
  \citenamefont {He},\ and\ \citenamefont {Shen}}]{r46}%
  \BibitemOpen
  \bibfield  {author} {\bibinfo {author} {\bibfnamefont {J.~A.}\ \bibnamefont
  {Sobota}}, \bibinfo {author} {\bibfnamefont {Y.}~\bibnamefont {He}},\ and\
  \bibinfo {author} {\bibfnamefont {Z.-X.}\ \bibnamefont {Shen}},\ }\href
  {https://doi.org/10.1103/RevModPhys.93.025006} {\bibfield  {journal}
  {\bibinfo  {journal} {Rev. Mod. Phys.}\ }\textbf {\bibinfo {volume} {93}},\
  \bibinfo {pages} {025006} (\bibinfo {year} {2021})}\BibitemShut {NoStop}%
\bibitem [{\citenamefont {Yi}\ \emph {et~al.}(2015)\citenamefont {Yi},
  \citenamefont {Wang}, \citenamefont {Kemper}, \citenamefont {Mo},
  \citenamefont {Hussain}, \citenamefont {Bourret-Courchesne}, \citenamefont
  {Lanzara}, \citenamefont {Hashimoto}, \citenamefont {Lu}, \citenamefont
  {Shen},\ and\ \citenamefont {Birgeneau}}]{r47}%
  \BibitemOpen
  \bibfield  {author} {\bibinfo {author} {\bibfnamefont {M.}~\bibnamefont
  {Yi}}, \bibinfo {author} {\bibfnamefont {M.}~\bibnamefont {Wang}}, \bibinfo
  {author} {\bibfnamefont {A.~F.}\ \bibnamefont {Kemper}}, \bibinfo {author}
  {\bibfnamefont {S.~K.}\ \bibnamefont {Mo}}, \bibinfo {author} {\bibfnamefont
  {Z.}~\bibnamefont {Hussain}}, \bibinfo {author} {\bibfnamefont
  {E.}~\bibnamefont {Bourret-Courchesne}}, \bibinfo {author} {\bibfnamefont
  {A.}~\bibnamefont {Lanzara}}, \bibinfo {author} {\bibfnamefont
  {M.}~\bibnamefont {Hashimoto}}, \bibinfo {author} {\bibfnamefont {D.~H.}\
  \bibnamefont {Lu}}, \bibinfo {author} {\bibfnamefont {Z.~X.}\ \bibnamefont
  {Shen}},\ and\ \bibinfo {author} {\bibfnamefont {R.~J.}\ \bibnamefont
  {Birgeneau}},\ }\href {https://doi.org/10.1103/PhysRevLett.115.256403}
  {\bibfield  {journal} {\bibinfo  {journal} {Phys. Rev. Lett.}\ }\textbf
  {\bibinfo {volume} {115}},\ \bibinfo {pages} {256403} (\bibinfo {year}
  {2015})}\BibitemShut {NoStop}%
\bibitem [{\citenamefont {Ueno}\ \emph {et~al.}(2011)\citenamefont {Ueno},
  \citenamefont {Nakamura}, \citenamefont {Shimotani}, \citenamefont {Yuan},
  \citenamefont {Kimura}, \citenamefont {Nojima}, \citenamefont {Aoki},
  \citenamefont {Iwasa},\ and\ \citenamefont {Kawasaki}}]{r52}%
  \BibitemOpen
  \bibfield  {author} {\bibinfo {author} {\bibfnamefont {K.}~\bibnamefont
  {Ueno}}, \bibinfo {author} {\bibfnamefont {S.}~\bibnamefont {Nakamura}},
  \bibinfo {author} {\bibfnamefont {H.}~\bibnamefont {Shimotani}}, \bibinfo
  {author} {\bibfnamefont {H.~T.}\ \bibnamefont {Yuan}}, \bibinfo {author}
  {\bibfnamefont {N.}~\bibnamefont {Kimura}}, \bibinfo {author} {\bibfnamefont
  {T.}~\bibnamefont {Nojima}}, \bibinfo {author} {\bibfnamefont
  {H.}~\bibnamefont {Aoki}}, \bibinfo {author} {\bibfnamefont {Y.}~\bibnamefont
  {Iwasa}},\ and\ \bibinfo {author} {\bibfnamefont {M.}~\bibnamefont
  {Kawasaki}},\ }\href {https://doi.org/10.1038/nnano.2011.78} {\bibfield
  {journal} {\bibinfo  {journal} {Nat. Nanotechnol.}\ }\textbf {\bibinfo
  {volume} {6}},\ \bibinfo {pages} {408} (\bibinfo {year} {2011})}\BibitemShut
  {NoStop}%
\bibitem [{\citenamefont {Ren}\ \emph {et~al.}(2022)\citenamefont {Ren},
  \citenamefont {Li}, \citenamefont {Sun}, \citenamefont {Ju}, \citenamefont
  {Liu}, \citenamefont {Hong}, \citenamefont {Sun}, \citenamefont {Tao},
  \citenamefont {Zhou}, \citenamefont {Xu},\ and\ \citenamefont {Xie}}]{r53}%
  \BibitemOpen
  \bibfield  {author} {\bibinfo {author} {\bibfnamefont {T.}~\bibnamefont
  {Ren}}, \bibinfo {author} {\bibfnamefont {M.}~\bibnamefont {Li}}, \bibinfo
  {author} {\bibfnamefont {X.}~\bibnamefont {Sun}}, \bibinfo {author}
  {\bibfnamefont {L.}~\bibnamefont {Ju}}, \bibinfo {author} {\bibfnamefont
  {Y.}~\bibnamefont {Liu}}, \bibinfo {author} {\bibfnamefont {S.}~\bibnamefont
  {Hong}}, \bibinfo {author} {\bibfnamefont {Y.}~\bibnamefont {Sun}}, \bibinfo
  {author} {\bibfnamefont {Q.}~\bibnamefont {Tao}}, \bibinfo {author}
  {\bibfnamefont {Y.}~\bibnamefont {Zhou}}, \bibinfo {author} {\bibfnamefont
  {Z.~A.}\ \bibnamefont {Xu}},\ and\ \bibinfo {author} {\bibfnamefont
  {Y.}~\bibnamefont {Xie}},\ }\href {https://doi.org/10.1126/sciadv.abn4273}
  {\bibfield  {journal} {\bibinfo  {journal} {Sci. Adv.}\ }\textbf {\bibinfo
  {volume} {8}},\ \bibinfo {pages} {eabn4273} (\bibinfo {year}
  {2022})}\BibitemShut {NoStop}%
\bibitem [{\citenamefont {Wemple}(1965)}]{r54}%
  \BibitemOpen
  \bibfield  {author} {\bibinfo {author} {\bibfnamefont {S.~H.}\ \bibnamefont
  {Wemple}},\ }\href {https://doi.org/10.1103/PhysRev.137.A1575} {\bibfield
  {journal} {\bibinfo  {journal} {Phys. Rev.}\ }\textbf {\bibinfo {volume}
  {137}},\ \bibinfo {pages} {A1575} (\bibinfo {year} {1965})}\BibitemShut
  {NoStop}%
\bibitem [{\citenamefont {Thompson}\ \emph {et~al.}(1982)\citenamefont
  {Thompson}, \citenamefont {Boatner},\ and\ \citenamefont {Thomson}}]{r55}%
  \BibitemOpen
  \bibfield  {author} {\bibinfo {author} {\bibfnamefont {J.~R.}\ \bibnamefont
  {Thompson}}, \bibinfo {author} {\bibfnamefont {L.~A.}\ \bibnamefont
  {Boatner}},\ and\ \bibinfo {author} {\bibfnamefont {J.~O.}\ \bibnamefont
  {Thomson}},\ }\href {https://doi.org/10.1007/bf00683987} {\bibfield
  {journal} {\bibinfo  {journal} {Journal of Low Temperature Physics}\ }\textbf
  {\bibinfo {volume} {47}},\ \bibinfo {pages} {467} (\bibinfo {year}
  {1982})}\BibitemShut {NoStop}%
\bibitem [{\citenamefont {Vogt}\ and\ \citenamefont {Uwe}(1984)}]{r56}%
  \BibitemOpen
  \bibfield  {author} {\bibinfo {author} {\bibfnamefont {H.}~\bibnamefont
  {Vogt}}\ and\ \bibinfo {author} {\bibfnamefont {H.}~\bibnamefont {Uwe}},\
  }\href {https://doi.org/10.1103/PhysRevB.29.1030} {\bibfield  {journal}
  {\bibinfo  {journal} {Phys. Rev. B}\ }\textbf {\bibinfo {volume} {29}},\
  \bibinfo {pages} {1030} (\bibinfo {year} {1984})}\BibitemShut {NoStop}%
\bibitem [{\citenamefont {Vogt}(1988)}]{r57}%
  \BibitemOpen
  \bibfield  {author} {\bibinfo {author} {\bibfnamefont {H.}~\bibnamefont
  {Vogt}},\ }\href {https://doi.org/10.1103/physrevb.38.5699} {\bibfield
  {journal} {\bibinfo  {journal} {Phys. Rev. B}\ }\textbf {\bibinfo {volume}
  {38}},\ \bibinfo {pages} {5699} (\bibinfo {year} {1988})}\BibitemShut
  {NoStop}%
\bibitem [{\citenamefont {Perry}\ \emph {et~al.}(1989)\citenamefont {Perry},
  \citenamefont {Currat}, \citenamefont {Buhay}, \citenamefont {Migoni},
  \citenamefont {Stirling},\ and\ \citenamefont {Axe}}]{r58}%
  \BibitemOpen
  \bibfield  {author} {\bibinfo {author} {\bibfnamefont {C.~H.}\ \bibnamefont
  {Perry}}, \bibinfo {author} {\bibfnamefont {R.}~\bibnamefont {Currat}},
  \bibinfo {author} {\bibfnamefont {H.}~\bibnamefont {Buhay}}, \bibinfo
  {author} {\bibfnamefont {R.~M.}\ \bibnamefont {Migoni}}, \bibinfo {author}
  {\bibfnamefont {W.~G.}\ \bibnamefont {Stirling}},\ and\ \bibinfo {author}
  {\bibfnamefont {J.~D.}\ \bibnamefont {Axe}},\ }\href
  {https://doi.org/10.1103/physrevb.39.8666} {\bibfield  {journal} {\bibinfo
  {journal} {Phys. Rev. B}\ }\textbf {\bibinfo {volume} {39}},\ \bibinfo
  {pages} {8666} (\bibinfo {year} {1989})}\BibitemShut {NoStop}%
\bibitem [{\citenamefont {Jandl}\ \emph {et~al.}(1991)\citenamefont {Jandl},
  \citenamefont {Banville}, \citenamefont {Dufour}, \citenamefont {Coulombe},\
  and\ \citenamefont {Boatner}}]{r59}%
  \BibitemOpen
  \bibfield  {author} {\bibinfo {author} {\bibfnamefont {S.}~\bibnamefont
  {Jandl}}, \bibinfo {author} {\bibfnamefont {M.}~\bibnamefont {Banville}},
  \bibinfo {author} {\bibfnamefont {P.}~\bibnamefont {Dufour}}, \bibinfo
  {author} {\bibfnamefont {S.}~\bibnamefont {Coulombe}},\ and\ \bibinfo
  {author} {\bibfnamefont {L.~A.}\ \bibnamefont {Boatner}},\ }\href
  {https://doi.org/10.1103/physrevb.43.7555} {\bibfield  {journal} {\bibinfo
  {journal} {Phys. Rev. B}\ }\textbf {\bibinfo {volume} {43}},\ \bibinfo
  {pages} {7555} (\bibinfo {year} {1991})}\BibitemShut {NoStop}%
\bibitem [{\citenamefont {Ibach}(1970)}]{r60}%
  \BibitemOpen
  \bibfield  {author} {\bibinfo {author} {\bibfnamefont {H.}~\bibnamefont
  {Ibach}},\ }\href {https://doi.org/10.1103/PhysRevLett.24.1416} {\bibfield
  {journal} {\bibinfo  {journal} {Phys. Rev. Lett.}\ }\textbf {\bibinfo
  {volume} {24}},\ \bibinfo {pages} {1416} (\bibinfo {year}
  {1970})}\BibitemShut {NoStop}%
\bibitem [{\citenamefont {Lourenço-Martins}\ and\ \citenamefont
  {Kociak}(2017)}]{r61}%
  \BibitemOpen
  \bibfield  {author} {\bibinfo {author} {\bibfnamefont {H.}~\bibnamefont
  {Lourenço-Martins}}\ and\ \bibinfo {author} {\bibfnamefont {M.}~\bibnamefont
  {Kociak}},\ }\href {https://doi.org/10.1103/PhysRevX.7.041059} {\bibfield
  {journal} {\bibinfo  {journal} {Phys. Rev. X}\ }\textbf {\bibinfo {volume}
  {7}},\ \bibinfo {pages} {041059} (\bibinfo {year} {2017})}\BibitemShut
  {NoStop}%
\bibitem [{\citenamefont {Fuchs}\ and\ \citenamefont {Kliewer}(1965)}]{r62}%
  \BibitemOpen
  \bibfield  {author} {\bibinfo {author} {\bibfnamefont {R.}~\bibnamefont
  {Fuchs}}\ and\ \bibinfo {author} {\bibfnamefont {K.~L.}\ \bibnamefont
  {Kliewer}},\ }\href {https://doi.org/10.1103/PhysRev.140.A2076} {\bibfield
  {journal} {\bibinfo  {journal} {Phys. Rev.}\ }\textbf {\bibinfo {volume}
  {140}},\ \bibinfo {pages} {A2076} (\bibinfo {year} {1965})}\BibitemShut
  {NoStop}%
\bibitem [{\citenamefont {Dubois}\ and\ \citenamefont {Schwartz}(1982)}]{r63}%
  \BibitemOpen
  \bibfield  {author} {\bibinfo {author} {\bibfnamefont {L.~H.}\ \bibnamefont
  {Dubois}}\ and\ \bibinfo {author} {\bibfnamefont {G.~P.}\ \bibnamefont
  {Schwartz}},\ }\href {https://doi.org/10.1103/PhysRevB.26.794} {\bibfield
  {journal} {\bibinfo  {journal} {Phys. Rev. B}\ }\textbf {\bibinfo {volume}
  {26}},\ \bibinfo {pages} {794} (\bibinfo {year} {1982})}\BibitemShut
  {NoStop}%
\bibitem [{\citenamefont {Mahan}(2000)}]{r64}%
  \BibitemOpen
  \bibfield  {author} {\bibinfo {author} {\bibfnamefont {G.~D.}\ \bibnamefont
  {Mahan}},\ }in\ \href@noop {} {\emph {\bibinfo {booktitle} {Many-Particle
  Physics}}}\ (\bibinfo  {publisher} {Plenum},\ \bibinfo {address} {New York},\
  \bibinfo {year} {2000})\ pp.\ \bibinfo {pages} {224--226},\ \bibinfo
  {edition} {3rd}\ ed.\BibitemShut {Stop}%
\bibitem [{\citenamefont {Marsiglio}(2020)}]{r65}%
  \BibitemOpen
  \bibfield  {author} {\bibinfo {author} {\bibfnamefont {F.}~\bibnamefont
  {Marsiglio}},\ }\href {https://doi.org/10.1016/j.aop.2020.168102} {\bibfield
  {journal} {\bibinfo  {journal} {Annals of Physics}\ }\textbf {\bibinfo
  {volume} {417}},\ \bibinfo {pages} {168102} (\bibinfo {year}
  {2020})}\BibitemShut {NoStop}%
\end{thebibliography}

%apsrev4-2.bst 2019-01-14 (MD) hand-edited version of apsrev4-1.bst
%Control: key (0)
%Control: author (8) initials jnrlst
%Control: editor formatted (1) identically to author
%Control: production of article title (0) allowed
%Control: page (0) single
%Control: year (1) truncated
%Control: production of eprint (0) enabled
%

\end{document}